\documentclass[aps, prd, twocolumn, showpacs, floatfix, letterpaper,
 nofootinbib, superscriptaddress,longbibliography]{revtex4-1}
\usepackage{graphicx}
\usepackage{epsfig}
\usepackage{bm}
\usepackage{amsfonts}
\usepackage{pgf}
\usepackage{color}
\usepackage[T1]{fontenc}
\usepackage[latin1]{inputenc}
\usepackage{graphicx}
\usepackage[english]{babel}
\usepackage{amsmath}
\usepackage{amssymb}
\usepackage{amsfonts}
\usepackage{makecell}
\usepackage{hyperref}
\usepackage[draft,deletedmarkup=xout]{changes}
\usepackage{mathtools}
\definecolor{green}{rgb}{0.19,0.64,0.54}
\definecolor{blue}{rgb}{0,0,1}
\definecolor{reddish}{rgb}{0.65, 0.2, 0.2}
\definecolor{darkgreen}{rgb}{0.2,0.7,0.3}
\definecolor{darkblue}{rgb}{0.3,0.40,0.48}
\definecolor{gray}{rgb}{.8,.8,.8}

\hypersetup{,
	pdftitle={EmptyBianchI},
  	pdfauthor={P. Malkiewicz et al.},
    colorlinks=true,       
    linkcolor=darkblue,          
    citecolor=darkgreen,        
    filecolor=reddish,      
    urlcolor=blue,           
    linktocpage=true
}

\newcommand{\dd}{\mathrm{d}}
\newcommand{\ex}{\mathrm{e}}
\newcommand{\GN}{G_\mathrm{_N}}

\newcommand{\Ka}{\mathcal{K}}
\newcommand\underrel[2]{\mathrel{\mathop{#2}\limits_{#1}}}
\usepackage{ulem}

\begin{document}
	
\title{Quantum empty Bianchi I spacetime with internal time}

\author{Przemys{\l}aw Ma{\l}kiewicz}
\email{Przemyslaw.Malkiewicz@ncbj.gov.pl}

\affiliation{National Centre for Nuclear Research, Pasteura 7, 
02-093 Warszawa, Poland}
	
\author{Patrick Peter} \email{peter@iap.fr}
\affiliation{${\cal G}\mathbb{R}\varepsilon\mathbb{C}{\cal O}$ -- Institut
d'Astrophysique de Paris, CNRS \& Sorbonne Universit\'e, UMR 7095
98 bis boulevard Arago, 75014 Paris, France}
\affiliation{Department of Applied Mathematics and Theoretical Physics,
Centre for Mathematical Sciences, University of Cambridge, Wilberforce
Road, Cambridge CB3 0WA, United Kingdom.}

\author{S. D. P. Vitenti} \email{vitenti@uel.br}
\affiliation{Departamento de F\'{i}sica, Universidade Estadual 
	de Londrina, Rod. Celso Garcia Cid, Km 380, 86057-970, 
	Londrina, Paran\~a, Brazil.}
\affiliation{Instituto de F\'{\i}sica - Universidade de Bras\'{\i}lia - UnB,
Campus Universit\'ario Darcy Ribeiro-Asa Norte Sala\\ BT 297-ICC-Centro,
70919-970 Bras\'{\i}lia, Brazil.}

\begin{abstract}

We discuss the question of time in a Bianchi I quantum cosmology in the
framework of singularity avoidance. We show that time parameters fall
into two distinct classes, that are such that the time development of the
wave function either always leads to the appearance of a singularity
(fast-gauge time) or that always prevents it from occurring (slow-gauge
time). Furthermore, we find that, in the latter case, there exists an
asymptotic regime, independent of the clock choice. This may point to a
possible solution of the clock issue in quantum cosmology if there exists
a suitable class of clocks all yielding identical relevant physical
consequences.

\end{abstract}

\date{\today}
\maketitle

\section*{Introduction}

The problem of time \cite{Isham:1992ms, Zeh:1992vf, Anderson:2012vk,
Kiefer:2009tq} in quantum gravity \cite{Kiefer2007,Kiefer:2013jqa} is a
longstanding one \cite{DeWitt:1967yk} that stems from the fact that the
underlying notions in general relativity (GR) and quantum theory are
incompatible. Among the numerous proposals that have been suggested is
that of using a perfect fluid \cite{Brown:1994py} whose Hamiltonian,
being linear in a momentum, naturally transforms the Wheeler-De Witt
equation in the Schr\"odinger form upon quantization of this momentum.
Such a solution also permits, in the trajectory approach of quantum
mechanics, to naturally avoid cosmological singularities
\cite{AcaciodeBarros1998, Peter2016b, Peter:2018knf}. Note that using an
internal degree of freedom to define time was also used in completely
different contexts; see, e.g., Ref.~\cite{mott1931}.

It is well known that the canonical formulation of general relativity is
given by a totally constrained Hamiltonian system. In these systems the
dynamics takes place inside the first-class constraints hypersurface and
as a consequence the symplectic structure projected on this hypersurface
is no longer symplectic (it is actually a presymplectic form). The
hypersurface orbits generated by the vector constraint representing
infinitesimal spatial diffeomorphisms are removed by imposing appropriate
gauge-fixing conditions, whereas the hypersurface orbits generated by the
Hamiltonian constraint representing the physical motion of the
gravitational system require another treatment.

For this physical motion, one can choose a foliation for the constrained
manifold such that in each sheet the restriction of the 2-form yields
again a symplectic structure, which consequently defines a Hamiltonian
system (where the Hamiltonian generates the motion across sheets). In
this language, the example of the perfect fluid cited above (with spatial
diffeomorphisms being absent), where the Hamiltonian depends linearly on
a momentum variable $p_T$ and is independent of its conjugate variable
$T$, defines a natural foliation: the constraint is solved in terms of
$p_T$ and  $T = \text{const.}$ gives the desired foliation and
consequently $p_T$ generates ``time'' translations. Thus, as discussed in
\cite{Rovelli:1990jm}, even at the classical level, if the constraints
hypersurface $\mathcal{M}$ does not admit a global foliation of the form
$\mathcal{M} = \mathbb{R}\times\mathcal{N}$, it seems that even
classically one cannot define time globally in these cases.

When a time variable is well defined (and consequently a foliation on the
constraints hypersurface), it is usually not unique. Classically, this is
not a problem: it just corresponds to different ways to parametrize time
with respect to some internal degree of freedom, using a combination of
the canonical variables. On the other hand, it is not clear if this
choice of time variable affects the quantized system, i.e., if the choice
of time variable implies any physical consequence for the quantized
system. Arguably, one can demand, as a desired property of a reasonable
quantum gravity theory, that changing the time variable should not modify
the physical content of the theory. However, within the reduced phase
space approach, where we use a particular foliation for $\mathcal{M}$,
considered in the present paper we explicitly show that this seemingly
reasonable property does not necessarily hold once the system is
quantized. Let us briefly overview the reason why this happens postponing
a more detailed description to the later sections.

The key observation to make here is that the gauge-invariant content of a
constrained theory, such as canonical general relativity, comprises
expressions solely in terms of gauge-invariant variables, defined as
quantities that commute with the constraints; they are called Dirac
observables. Since the Hamiltonian in canonical relativity is a
constraint itself, no dynamical variable can be gauge invariant as none
can commute with the Hamiltonian constraint. As a result, dynamical
variables are not equipped with a well-defined Poisson structure, and
this forbids straightforward quantization. As discussed above, one way to
circumvent this difficulty is to introduce a foliation with an internal
time variable which, by assumption, commutes with all the Dirac
observables. This way, the internal time and all the dynamical
quantities, which are functions of Dirac observables, can be included in
the reduced phase space formalism. An immediate consequence of this
prescription is that the commutation relations involving dynamical
variables depend on the choice of internal time.

Another idea to circumvent the conceptual difficulty discussed above is
expressed in the idea of evolving constants \cite{Rovelli:1990jm,
Rovelli:1989jn, Dittrich:2005kc}. This approach relies on the observation
that Dirac observables, though nondynamical themselves, could in fact be
seen as particular functions of dynamical variables and therefore
encoding the dynamics as relations between these variables. For instance,
the value of a dynamical quantity $X$ when another dynamical quantity $Y$
takes the value $y=y_0$ is constant along any dynamical trajectory and is
thus a Dirac observable. In this way, the entire dynamical trajectory of
$X$ is given by a $y$-parametrized family of Dirac observables. As far as
we can tell, such an approach does not contradict that presented below.
We note that the values of a dynamical quantity $X$ when another
dynamical quantity $Z$ takes the values $z=z_0$ form a different, now
$z$-parametrized, family of Dirac observables. Hence, for a fixed
quantization of Dirac observables, the $y$ and $z$families may exhibit
different quantum properties though they describe the dynamics of the
same dynamical observable, $X$. Therefore, it is meaningful to speak
about dynamical observables at the quantum level only with respect to
specific internal time variables.

In this paper, we discuss the quantization of the vacuum Bianchi I case,
showing how it generates a time whose arbitrariness in the definition
produces a clock-choice issue. We discuss some choices (fast and slow
gauge times), and this leads to a possible criterion: some clocks, upon
quantization of the system, are singularity free, while others do exhibit
a singularity. By imposing a specific ordering of the operators in the
Hamiltonian, we can put the latter in a canonical form and obtain exact
singularity-free solutions for the average trajectories. We provide a
clear illustration of the dependence of quantum dynamics on the choice of
internal time. Surprisingly, we identify a certain property of quantum
gravitational dynamics which does not depend on the choice of internal
time and points to a possible solution of the time problem.

\section{Empty Bianchi I}
\label{Model}

Our starting point is the vacuum GR gravitation theory, whose
classical Einstein-Hilbert action $\mathcal{S}$ reads, in units
with $8\pi\GN=1$,
\begin{equation}
\mathcal{S} = \frac{1}{2} \int R\sqrt{-g}\,\dd^4x.
\label{SEH}
\end{equation}
This theory admits the Bianchi I metric, given in terms of the
lapse function $N$ by
\begin{equation}\label{eq:metric}
\dd s^2 = -N^2\dd \tau^2 + \sum_{i=1}^3 a_i^2\left(\dd x^i\right)^2,
\end{equation}
as a solution of the corresponding vacuum Einstein equations for
a flat homogeneous but anisotropic spacetime. 

The scale factors associated to each direction can be recast as \cite{wainwright_ellis_1997}
\begin{align}\label{a1_to_beta}
a_1 &= \ex^{\beta_0 + \beta_+ + \sqrt{3}\beta_-},\\ 
a_2 &= \ex^{\beta_0 + \beta_+ - \sqrt{3}\beta_-},\\
a_3 &= \ex^{\beta_0 - 2\beta_+},
\end{align}
where we introduced the anisotropy variables $\beta_\pm$ and
$\beta_0$, the latter providing the volume $V$ of the manifold, assumed
compact, through
\begin{equation}\label{volume}
V \equiv a_1a_2a_3 = \ex^{3\beta_0}.
\end{equation}

The action $\mathcal{S}$ for the metric \eqref{eq:metric} reads
\begin{equation}\label{action}
\mathcal{S} = \int\dd \tau \left(p_0\dot{\beta}_0 + p_+\dot{\beta}_+ +
p_-\dot{\beta}_- - N C\right),
\end{equation}
where the Hamiltonian $H = N C$ is such that the constraint $C$ satisfies
\begin{equation}\label{hamiltonian}
C = \frac{\ex^{-3\beta_0}}{24}\left(-p_0^2 + p_+^2 + p_-^2\right).
\end{equation}
The canonical one-form can be read directly from Eq.~\eqref{action} as
\begin{equation}\label{coform}
\dd\theta = p_0\dd\beta_0 + p_+\dd\beta_+ + p_-\dd\beta_-.
\end{equation}
In terms of this one form the action is
\begin{equation}\label{action2}
\mathcal{S} = \int\dd \tau \left(\frac{\dd\theta}{\dd\tau} - NC\right).
\end{equation}

The volume variable $V$ turns out to be more convenient than $\beta_0$.
One has
\begin{equation}\label{dbeta}
\dd\beta_0 = \frac{\ex^{-3\beta_0}}{3}\dd V,
\end{equation}
and the new momentum associated to it has to be 
\begin{equation}\label{pv}
p_V \equiv \frac{\ex^{-3\beta_0}}{3} p_0,
\end{equation}
in order to keep the one-form canonical, i.e.,
\begin{equation}
\dd\theta = p_V\dd V + p_+\dd\beta_+ + p_-\dd\beta_-.
\end{equation}
Using this new variable the constraint (\ref{hamiltonian}) is written as
\begin{equation}
C = \frac{3V}{8}\left(-p_V^2 + \frac{p_+^2 + p_-^2}{9V^2}\right).
\end{equation}
This constrained system must classically satisfy
\begin{equation}
C = 0,
\end{equation}
the quantization of which we turn to below.

Let us first parametrize the above problem explicitly, and to achieve
that goal first rewrite the problem using variables that evince the
system symmetries. The variables $\beta_\pm$ are clearly cyclic, and
therefore their momenta are conserved, i.e.,
\begin{equation}
\dot{p}_\pm = 0. 
\end{equation}
To avoid carrying these two constants around we perform the
transformation
\begin{equation}\label{t1}
p_+ = k \cos\alpha, \quad p_- = k \sin\alpha,
\end{equation} 
where we can choose $k>0$ without loss of generality. Ensuring the
one-form remains canonical, we obtain
\begin{align}\label{ntheta}
\dd\theta= p_V\dd V + p_k\dd k + p_\alpha\dd\alpha+\textsc{s.t.},
\end{align}
where we defined the new two momentum variables
\begin{align}
p_k &\equiv - \left(\cos\alpha\beta_+ + \sin\alpha\beta_-\right), \\
p_\alpha &\equiv\left(k\sin\alpha\beta_+ - k\cos\alpha\beta_-\right),
\end{align}
and the surface term $\textsc{s.t.} =\dd\left(k \cos\alpha\beta_+ + k
\sin\alpha\beta_-\right)$ in Eq.~\eqref{ntheta} is an exact form, which
we can and thus will ignore from here on. Note also that neither
$p_\alpha$ nor $\alpha$ appears in the Hamiltonian and consequently both
are constant. We shall thus also ignore them.

In terms of the above variables, our system is described by the
action~\eqref{action2}, where the canonical one-form and the
constraint are
\begin{align}
\dd\theta &= p_V\dd V + p_k\dd k, \\
C &= \frac{3V}{8}\left(-p_V^2 + \frac{k^2}{9V^2}\right).
\end{align}

\section{Parametrizing the problem}

The system action~\eqref{action2} is constrained. The lapse function $N$
acts as a Lagrange multiplier and imposes that $C = 0$. It turns out that
one can solve this constraint explicitly and then obtain a parametrized
Hamiltonian. While this is a trivial recasting of the classical problem,
when we move to quantization this has a nontrivial effect. The
parametrization of the problem involves turning one of its degrees of
freedom in a monotonically evolving variable which, upon quantization,
acts as a time in the corresponding Schr\"{o}dinger equation. It
therefore acquires a different status than the other variables: with a
physical clock (which in our case is internal to the system) thus
defined, this entails the existence of a time parameter related to that
particular clock, in terms of which one derives the evolution of the
dynamical variables.

Before starting with the parametrization, it is useful to study the
Hamilton equations of motion of our problem. They read
\begin{equation}
\begin{array}{rl}
\dot{k} &= 0, \\
\dot{p}_k &= -\displaystyle\frac{k}{12V} N , \\
&\\
\dot{V} &= -\displaystyle\frac{3Vp_V}{4} N, \\
\dot{p}_V &= -\left[\displaystyle\frac{3}{8}\left(-p_V^2 +
\frac{k^2}{9V^2}\right) - \frac{k^2}{12V^2}\right] N,
\end{array}
\label{Ham}
\end{equation}
together with the constraint
\begin{equation}\label{c1}
\frac{3V}{8}\left(-p_V^2 + \frac{k^2}{9V^2}\right) = 0.
\end{equation}
Since $V \neq 0$, Eqs.~\eqref{Ham} reduce to
\begin{equation}
\dot{k} = 0 \ \ \ \hbox{and}\ \ \ \dot{V} = -\frac{3Vp_V}{4}N,
\label{eqm11}
\end{equation}
for the variables, and
\begin{equation}
\dot{p}_k = -\frac{k}{12V}N  \ \ \ \hbox{and}\ \ \ \dot{p}_V =
\frac{3p_V^2}{4}N \label{eqm12}
\end{equation}
for the associated momenta.

The system above is closed for $p_V$ and $V$ and therefore can be solved
first for these two variables and then for $k$ and $p_k$ (when one has
to impose the constraint above when choosing the initial conditions for
$k$ and $p_k$).

\subsection{Reduced phase space and choice of time}\label{sec:reduced}

Thus far, we have not chosen the time variable $\tau$ appearing in the
line element \eqref{eq:metric}, and indeed the above problem can be
solved for any choice of this time, and hence of the lapse function $N$.
Indeed, in the previous section, we wrote the equations of motion as
derived from the Hamiltonian as first order in time, which we called
``$\tau$'' but otherwise left undefined, merely assuming there exists
such an ordering of events labeling. In order to move forward, we need to
be more specific in the choice of this time variable.

Classically, one can define/choose a time parametrization by solving the
constraint directly in the one-form $\dd\theta$: using $k^2=9 p_V^2 V^2$
(note that we do not have an ambiguity in choosing the sign of $k$ since
we have assumed $k>0$), we obtain
\begin{equation}
\dd\theta = p_V\dd V + \frac{p_k}{2k}\dd k^2 
= p_V\dd V + \frac{9p_k}{2k}\dd\left(V^2p_V^2\right).
\label{dtheta}
\end{equation}
Now, one can easily reduce the one-form above to a single term,
\begin{equation}
\dd\theta = \left(\frac{9p_k}{k}-
\frac{\ln V}{Vp_V}\right)\dd\left(\frac{V^2p_V^2}{2}\right)+\textsc{s.t.},
\end{equation}
and ignoring the surface term $\textsc{s.t.} =\dd\left(Vp_V\ln V\right)$
since it does not contribute to the action, we get
\begin{equation}
\dd\theta =-\frac{V^2p_V^2}{2} \dd \Upsilon,\label{dthdA}
\end{equation}
where we removed another surface term $\dd(\Upsilon V^2p_V^2/2)$, and
set
\begin{equation}
\Upsilon \equiv \frac{9p_k}{k}-\frac{\ln V}{Vp_V};
\label{defA}
\end{equation}
both $\Upsilon$ and $(Vp_V)$ are constants of the motion. 

Let us introduce an arbitrary function of the dynamical variables
$T(V,p_V,p_k)$, through which we define a time $t$
\begin{equation}
t=\Upsilon+T (V,p_V,p_k),
\label{NewVarsT}
\end{equation}
which also thus depends on the dynamical variables. Setting
\begin{eqnarray}
Q & \equiv Vp_V T, \label{NewVarsQ}
\\
p_Q & \equiv V p_V,
\label{NewVarsP}
\end{eqnarray}
and plugging \eqref{NewVarsQ} and \eqref{NewVarsP} into \eqref{dthdA}, we
get
\begin{equation}
\dd\theta = P_Q\dd Q - \frac{p_Q^2}{2}\dd t,
\end{equation}
where we again removed a surface term $-\frac{1}{2}\dd(
p_QQ)$. We note that the
role of the phase space function $T$ is twofold: it defines both the
time parameter $t$ and the position variable $Q$.

A given choice of $T$ therefore implies, once the equations of motion are
solved, a classical solution $Q(t)$. Assuming one can invert this
relation, one can thus find the interval over which the corresponding
time parameter varies. As the dynamics of the system is that of a freely
moving particle independently of the choice of $T$, the ranges of $Q$ and
$t$ must be related. Many cases are then possible, depending on whether
$Q$ and $t$ are bounded or unbounded. If the range of $Q$ is real
($Q\in\mathbb{R}$), then the motion is unbounded and the singularity is
never reached. If, on the other hand, the range of $Q$ contains a finite
limit, say $Q\in [Q_0,\infty[$ for instance, then the motion
originates/terminates at $Q=Q_0$ in a finite time and the dynamics is
singular. Indeed, from \eqref{NewVarsQ} and \eqref{NewVarsP} we obtain
the singularity time as $t_0=\Upsilon+Q_0/p_Q$. The former case is dubbed
the fast-gauge time because the relevant clock ticks an infinite number
of times before reaching the singularity, whereas the latter is known as
the slow-gauge time. We see below examples of both situations. In both
cases the time variable $t$ is globally well defined in the sense that it
is always growing as the particle approaches or recedes from the boundary
(or, minus infinity). We notice that any given value of the fast-gauge
clock is taken twice: once for the expanding and once for the contracting
universe, whereas any given value of the slow-gauge clock is taken only
once: either in the expanding or contracting universe, depending on
whether $t>t_0$ or $t<t_0$. As we see, this property of slow-gauge clocks
enables one to remove the singularity by quantization while extending the
time variable across $t_0$. In order to have a single point $t_0$ of the
time axis correspond to the singularity, we choose $T$ such that
$Q_0=0$.\footnote{Otherwise we would have a whole interval of values of
$t$ for which the system is singular.}

\subsection{Fast-gauge time $\tau$}

Let us first consider a fast-gauge time example and assume that
\begin{equation}
T_\mathrm{fast} = \frac{\ln V}{Vp_V},
\label{Tfast}
\end{equation}
which diverges for $V\to 0$. From \eqref{NewVarsQ}, we see that
the relevant canonical variable is $Q_\mathrm{fast} =  \ln V$ for a time
defined through \eqref{NewVarsT}, namely $t_\mathrm{fast} = 9p_k/k$,
which is indeed monotonically related to the original time. We expand
below on the properties of this choice.

\subsubsection{Classical time choice}

We begin by noting that it is possible to rewrite the equation for $p_k$
as
\begin{equation}\label{a1}
\frac{\dd}{\dd\tau}\left(9\frac{p_k}{k}\right) = -\frac{3}{4}\frac{N}{V},
\end{equation}
implying, as stated above, that the quantity $9p_k/k$ is a monotonic
function of the arbitrary time $\tau$ appearing in the metric
\eqref{eq:metric}, as $V$ is positive definite and $N$ is nonvanishing,
and hence either always positive or always negative. As a result, the
quantity $9p_k/k$ can itself be used as a time parameter. Assuming this
is the case, we choose $\tau = t_\mathrm{fast} = 9p_k/k$, which agrees
with our general framework \eqref{NewVarsT} with the fast-gauge time
function $T_\mathrm{fast}$ from \eqref{Tfast}, leading to the following
lapse function
\begin{equation}\label{a2}
N = -\frac{4}{3}V < 0.
\end{equation}
This clearly shows that this choice of time parameter is globally well
defined.

For the sake of clarity, we repeat below the steps of
Sec.~\ref{sec:reduced}, starting directly with the action. As we have
seen above, solving the constraint directly in the one-form $\dd\theta$
using $k^2=9 p_V^2 V^2$ leads to \eqref{dtheta}, and therefore to
\begin{equation}
\dd\theta= p_V\dd V + \dd\left(\frac{9p_k}{2k}V^2p_V^2\right) -
\frac{V^2p_V^2}{2}\dd\left(\frac{9p_k}{k}\right).
\end{equation}
We can again safely ignore the exact form above since it will not
contribute to the equations of motion. Since we solved the
constraint the action is now merely given by
\begin{equation}\label{actionf}
\mathcal{S} = \int \dd\theta = \int\dd\tau \left(p_V\dot V -
\frac{V^2p_V^2}{2}\right),
\end{equation}
where we simply relabeled $\tau \equiv 9p_k / k$. This is an
unconstrained one-dimensional system whose dynamics stems from the
Hamiltonian $H = V^2p_V^2/2$.

It is now a simple matter to check the Hamilton equations are indeed
those obtained earlier. Indeed, they read
\begin{equation}\label{1eom}
\dot{V} = V^2p_V, \qquad \dot{p}_V = -Vp_V^2,
\end{equation}
which are the  correct equations of motion after substitution of  the
lapse~\eqref{a2} in Eqs.~\eqref{eqm11} and \eqref{eqm12}. Once the
equations for $V$ and $p_V$ are solved, we can use the
constraint~\eqref{c1} to obtain $k^2$. Finally, since $\tau = 9p_k/k$,
we can determine both quantities
\begin{equation}
k = 3V\vert p_V\vert, \qquad \hbox{and} \qquad p_k = \frac{k\tau}{9}.
\end{equation}

We note that in the proposed internal time $\tau = 9p_k/k$, the
classical dynamics is completely determined as the solution to Eqs
(\ref{1eom}) reads
\begin{equation}
\frac{\dd}{\dd\tau}\left(V p_V\right) = 0 \qquad \Longrightarrow
\qquad Vp_V=V_0p_{V0},
\label{VpV}
\end{equation}
and
\begin{align}
V=V_0 \ex^{(Vp_V) \tau}\qquad \hbox{and} \qquad p_V=p_{V0}
\ex^{-(Vp_V)\cdot\tau}.
\end{align} 
The singularity is pushed to $\tau\rightarrow \pm\infty$ for expanding
and contracting universes, respectively. These sorts of internal times
are sometimes called fast-gauge times, while the slow-gauge times are
those in which the dynamics terminates at finite values.  It has been
conjectured \cite{Gotay1983} that the canonical quantization cannot
resolve the singularity problem in fast-gauge times since the Hamiltonian
flow is complete in this case. Although the relation between the
singularity resolution and the choice of time might be a more subtle
issue \cite{Lemos:1995bs}, it seems to us that using the fast-gauge
internal time chosen above can indeed not prevent the appearance of a
singularity even in the quantum case. Let us illustrate this point.

\subsubsection{Quantum dynamics}

\begin{figure*}[t]
\includegraphics[width=0.45\textwidth]{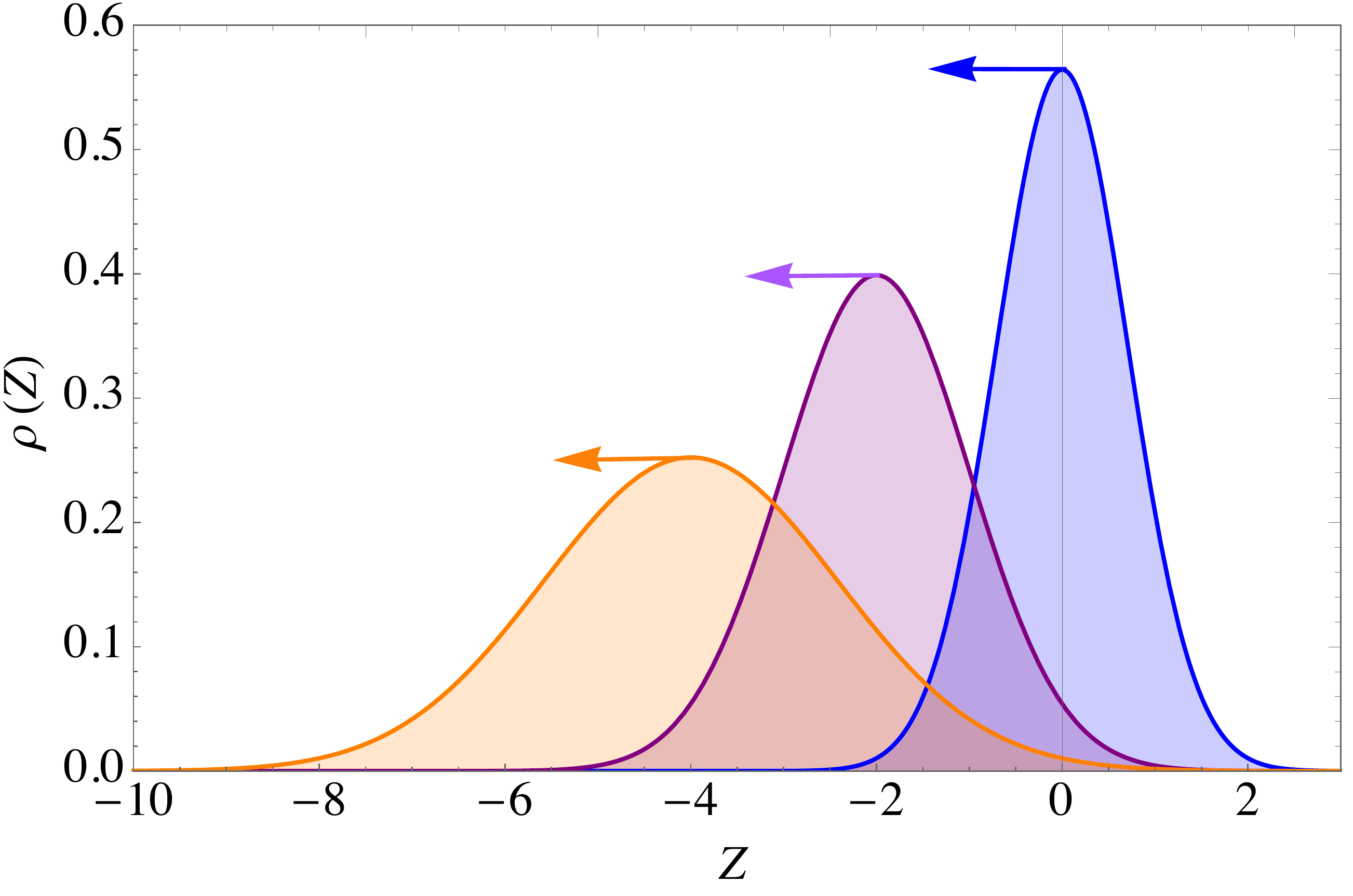}\null\hfill
\includegraphics[width=0.45\textwidth]{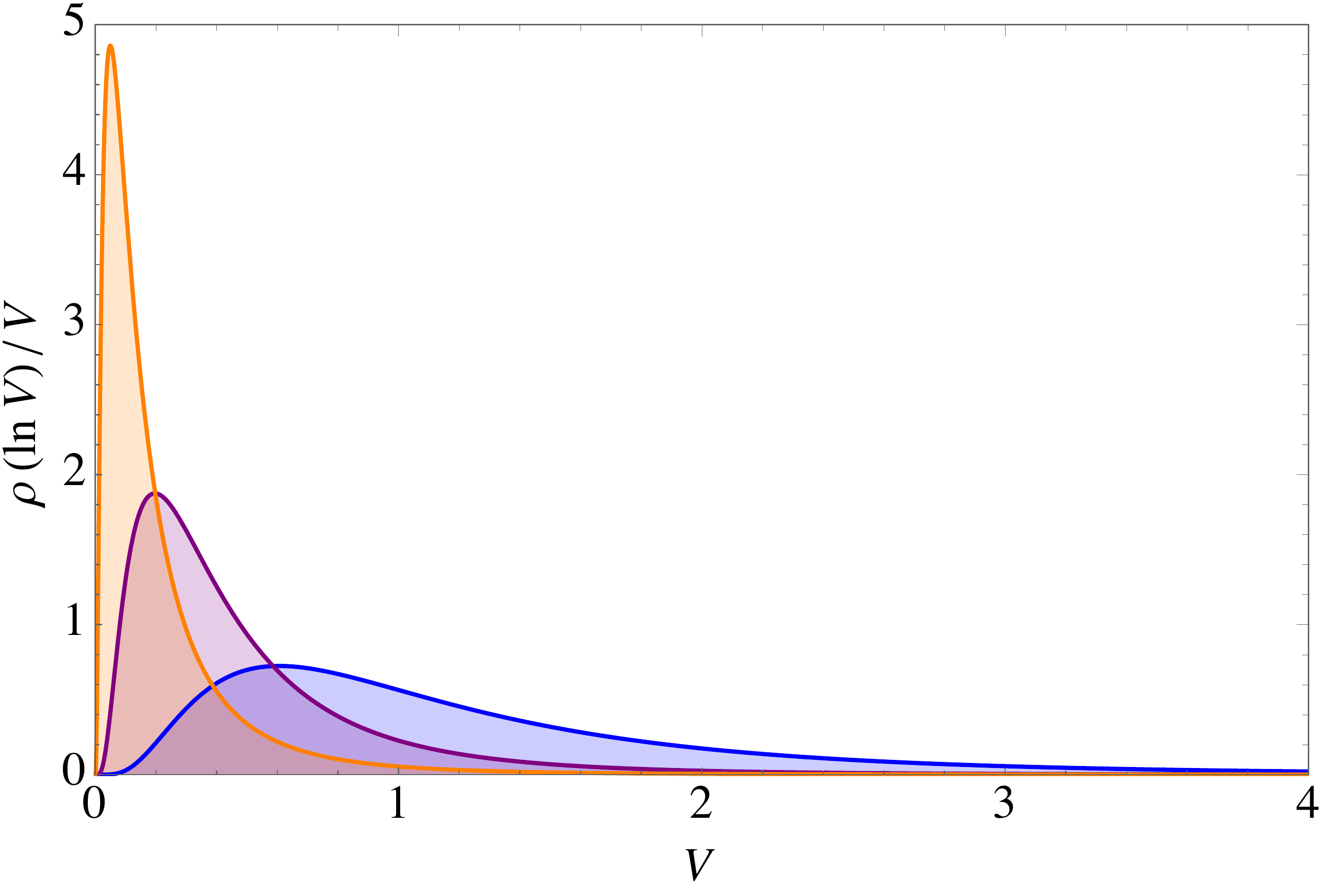}

\caption{The fast-gauge evolution of the probability distribution for a
Gaussian packet on the real line $Z$ (left panel) for $k=-1$,
$\tau=0,2,4$ and the respective packet to the half-line $V$ (right
panel) for $k=-1$ and $\tau=0,1,2$. Because of the packet spreading, the
probability density can initially grow with time for any sufficiently
large $V$. Nevertheless, the probability of finding the system on the
interval $[0,\epsilon ]$ for any $\epsilon>0$ tends to $1$ as
$\tau\rightarrow \infty$ and thus, the distribution converges to
$\delta(0)$. Notice that for every value of $\tau$ the probability
vanishes at $V=0$.}

\label{FG}
\end{figure*}

The Hamiltonian derived from the action (\ref{actionf}) and acting on the
half-plane phase space $(V,p_V)\in\mathbb{R}_+\times\mathbb{R}$ can be
promoted to a symmetric operator on a suitable dense subspace of the
Hilbert space of square-integrable functions on the half-line,
$L^2(\mathbb{R}_+,\dd V)$\footnote{The measure $\dd V$ is chosen so that
$\hat{V}=V$ and $\hat{p}_V=-i \partial_V$ are symmetric operators
provided the wave functions vanish both at $V=0$ and at $V\to \infty$. 
One could include an arbitrary function of the volume in the definition
of the measure such as, e.g., the scale factor $a =V^{1/3}$ and
correspondingly modify the definition of the relevant operators; such a
choice would however merely complicate matters with no physically
meaningful difference.}. One can choose the symmetric ordering
\begin{align}\label{FTH}
H=\frac12 V^2p_V^2\quad\mapsto \quad
\hat{H}=\frac12\sqrt{V}\frac{1}{i}\partial_V\sqrt{V}\cdot
\sqrt{V}\frac{1}{i}\partial_V\sqrt{V}.
\end{align}
In order to understand the quantum dynamics generated by the above
Hamiltonian, we make a coordinate transformation from the half-line to
the real line, $V\mapsto Q_\mathrm{fast} = \ln V \equiv Z$. The
corresponding unitary map between the respective Hilbert spaces,
$U:~L^2(\mathbb{R}_+,\dd V)\mapsto L^2(\mathbb{R},\dd Z)$, reads
\begin{widetext}
\begin{align}
\int_{\mathbb{R}_+} |\psi(V)|^2 \dd V = \int_{\mathbb{R}} 
|(U\psi)(Z)|^2 \dd Z \Longrightarrow \psi(V)\mapsto 
\left(U\psi\right)(Z)=\ex^{\frac{Z}{2}}\psi\left(\ex^Z\right).
\end{align}
It is straightforward to find that
\begin{align}
\left(\sqrt{V}\frac{1}{i}\partial_V\sqrt{V}\right)\,\psi(V)=
\ex^{-\frac{Z}{2}}\frac{1}{i}\partial_Z
\ex^{\frac{Z}{2}}\psi\left(\ex^Z\right),
 \end{align}
leading to
\begin{align}
U\left(\sqrt{V}\frac{1}{i}\partial_V\sqrt{V}\right)U^{-1}=\frac{1}{i}\partial_Z,
\end{align}
and hence
\begin{align}
\int_{\mathbb{R}_+} \psi^*(V) H(V)\psi(V) \dd V = \int_{\mathbb{R}}
|(U\psi)(Z)|^* \left(UHU^\dagger\right) \left( U\psi\right)(Z) \dd Z
\qquad \Longrightarrow\qquad
U\hat{H}U^{-1}=-\frac12 \partial_Z^2,\quad \hbox{and} \quad Z\in\mathbb{R}.
\end{align}
\end{widetext}
It is now clear that the Hamiltonian \eqref{FTH} must be essentially
self-adjoint and the unique dynamics it generates is unbounded with
wavepackets approaching the singularity $Z\rightarrow -\infty$ (i.e.,
$V\rightarrow 0$) as $\tau\rightarrow\pm\infty$, depending on the initial
condition. Figure \ref{FG} illustrates the fast-gauge evolution of the
probability distribution $\rho(Z,\tau)\equiv |\psi(Z,\tau)|^2$ carried by
a Gaussian wave packet as it approaches the singularity, $Z\rightarrow
-\infty$,
\begin{align}
\rho(Z,\tau)=\frac{1}{\sqrt{\pi \left( 1+\tau^2/4\right)}}
\exp\left[ -\displaystyle\frac{(Z-k\tau)^2}{1+\tau^2/4} \right],
\end{align}
which when mapped onto the half-line reads $\rho(\ln V,\tau)/V$ and
approaches the Dirac delta picked at $V=0$.  As the singularity does not
seem to be avoided in the present case let us now turn to considering a
slow-gauge internal time.

\subsection{Slow-gauge time $\eta$}

Let us now consider another transformation, using the function
\begin{equation}
T_\mathrm{slow} = \frac{1}{p_V},
\label{Tslow}
\end{equation}
whose limit is well defined when $V\to 0$: since $Vp_V=p_Q$ is a
constant, we must have $p_V \to \pm\infty$, and thus $T_\mathrm{slow} \to
0^\pm$. The choice \eqref{Tslow} translates into $t_\mathrm{slow} =
\Upsilon + 1/p_V$ and $Q_\mathrm{slow} = V$, again monotonically related
to the original time. Now $t_\mathrm{slow}$ is not defined in the full
real line, but only in two separate branches, namely $t_\mathrm{slow}\in
[\Upsilon, \infty[$ if $p_V>0$ or $t_\mathrm{slow}\in]-\infty, \Upsilon]$
for $p_V<0$. This entails a contracting universe ending at a singularity,
or an expanding one originating from a singularity. The complete solution
is then given by
\begin{equation}\label{key}
V = p_Q (t_\mathrm{slow} - \Upsilon).
\end{equation}
Let us develop these points.

As before, we solve the constraint directly in the one-form $\dd\theta$, 
now using a different parametrization, namely
\begin{equation}\label{slowgauge}\begin{split}
\dd\theta = & (Vp_V)\dd V
-\left(\frac{V^2p_V^2}{2}\right)\dd\left( \frac{9p_k}{k}+ \frac{V-\ln
V}{Vp_V}\right)\\
&+ \dd\left( \frac{9p_k}{2k}V^2p_V^2 +\frac{1}{2}V\ln V
 p_V-\frac{1}{2}V^2p_V\right).\end{split}
\end{equation}
We can again safely ignore the exact form above since it does  not
contribute to the equations of motion. Then, since we solved the
constraint the action is given by
\begin{equation}
\mathcal{S} = \int\dd\theta = \int\dd\eta \left(Vp_V\acute V -
\frac12 V^2p_V^2\right), \label{actionslow}
\end{equation}
where we  introduced the notation $\acute V\equiv\dd V/\dd\eta$ and
simply relabeled the new time variable $\eta$ through
\begin{equation}
\eta \equiv \frac{9p_k}{k}+ \frac{V-\ln V}{Vp_V} = t_\mathrm{slow},
\label{eta}
\end{equation}
which one can directly check indeed satisfies the requirements for being
a time, in the sense that it is a monotonic function: using the
equations of motion \eqref{eqm11} and \eqref{eqm12}, one readily obtains
\begin{equation}
\frac{\dd}{\dd\tau}\left(\frac{9p_k}{k}+ \frac{V-\ln V}{Vp_V}\right) = -\frac{3}{4}N.
\end{equation}
Note that the unconstrained Hamiltonian again is just $H =\frac12
V^2p_V^2$. However, unlike in the previous case, Eq.~\eqref{actionslow}
shows that it is now $p_VV$ and not $p_V$ anymore that plays the role of
the canonically conjugate momentum to the volume $V$. This seemingly
innocuous fact actually drastically transforms the problem as upon
introducing a new canonical variable, $\pi_V= p_Q =p_VV$, the Hamiltonian
$H$ again becomes that of a freely moving particle, but in this case the
dynamics is limited to the half-line
\begin{equation}
H=\frac{1}{2}\pi_V^2, \ \{V,\pi_V\}=1, \
\hbox{where}\  (V,\pi_V)\in\mathbb{R}_+\times\mathbb{R}.
\end{equation}
The dynamics therefore terminates at a finite value of $\eta$,
forwards/backwards in time for contracting/expanding universes,
respectively. As we show in the next section, in this case the
singularity can be resolved by quantization of the Hamiltonian
formalism.

\subsection{Other time variables $\eta'$}\label{clock_sec}

It is worth noting that there are many more allowed choices of time
variable when we parametrize the system. Let us consider a new internal
time,
\begin{align}\label{neweta}
\eta'=\eta'(\eta,V,\pi_V),
\end{align}
and redefine the dynamical variables,
\begin{align}\label{newcan}
\pi_V'=\pi_V,~~V'=V+\pi_V(\eta'-\eta).
\end{align}
Then Eq.~\eqref{slowgauge} without the exact form is
\begin{equation}
\dd\theta = \pi_V\dd V - \frac{\pi_V^2}{2}\dd\eta
= \pi_V'\dd V' - \frac{\pi_V'^2}{2}\dd\eta'+\dd\left[ \left( \eta-\eta' \right) 
\frac{\pi_V'^2}{2}\right].
\end{equation}
Since the exact form can be again ignored, the last expression above
shows that the formulation of the dynamics in a new internal time
(\ref{neweta}) is {\it formally} identical to the initial formulation
provided that Eq. (\ref{newcan}) holds. This property has significant
practical value as now it suffices to quantize {\it one} formalism in
order to obtain quantum formulation in {\it any} internal time
remembering that the basic variables may have different physical meaning
for different choices of time.

Note that the general transformation \eqref{neweta} and \eqref{newcan}
includes transformations to fast-gauge clocks, the situation that we want
to avoid. Indeed, writing the difference between the new and old time
variables, thereby defining the delay function $\Delta=\eta'-\eta$ from
now on, we find that one goes from the slow to the fast-gauge times
through
\begin{equation}
\Delta_{\mathrm{slow}\to \mathrm{fast}} = \frac{V-\ln V}{Vp_V},
\label{FastSlowTT}
\end{equation}
whose limit diverges when $V\to 0$. In order to ensure that such a
situation never occurs, we assume the transformation does not alter the
ranges of basic variables, i.e. we demand that
\begin{align}
0<V'=V+\pi_V \Delta <\infty.
\end{align}
If we furthermore assume that the delay function depends on the phase
space only, i.e.
\begin{align}
\Delta=\Delta(V,\pi_V),
\end{align}
the transformation \eqref{newcan} does not involve time variables. This
largely simplifies comparison between different time variables dynamics. 
Note that the new time variable is monotonic if and only if
\begin{align}
\frac{\dd\eta'}{\dd\eta}=\frac{\partial\eta'}{\partial\eta}+\{\Delta,H\} =
\frac{\partial\eta'}{\partial\eta}+ \left\{\Delta,\frac{1}{2}\pi_V^2\right\}
\neq 0,
\end{align}
i.e.,
\begin{equation}
\frac{\partial\Delta}{\partial\eta} + 1 + \pi_V \frac{\partial \Delta}{\partial V}
\neq 0,
\label{etaprimeOK}
\end{equation}
which in the simpler case $\partial\Delta/\partial\eta=0$ yields
\begin{align}
\frac{\partial V'}{\partial V}\neq 0.
\end{align}
Observe that this condition is equivalent to simply assuming that the
time transformation \eqref{neweta} and \eqref{newcan} is
$C^1$-invertible, ensuring that the canonical one-form $\dd\theta$ in
both parametrizations is identical (up to a total derivative).

\section{Quantization in the slow-time gauge}

Quantization of the half-plane phase space
$(V,\pi_V)\in\mathbb{R}_+\times\mathbb{R}$ is not an obvious
task.\footnote{Here and in what follows, we made a further canonical
transformation, namely $\pi_V \to \sqrt2\pi_V$ and $V \to V/\sqrt2$,
thus, removing the factor of one half from the Hamiltonian.} The problem
occurs because $\pi_V$ does not generate a global translation on that
phase space and the respective operator, $-i\partial_V$ on the half-line
$V>0$, admits no self-adjoint extension. Nevertheless, the square of this
operator, i.e. the (minus) Laplacian, can be given a self-adjoint
extension (in fact, it admits unaccountably infinite many such
extensions).

There are many ways to obtain a unitary evolution with the Laplacian, and
Sec.~\ref{quantize_sec} emphasizes one involving reordering of the basic
operators, making use of the commutation relations to produce a
``naturally'' self-adjoint Hamiltonian. Another, perhaps more
straightforward, method, which we discuss below, consists in restricting
the action of the Laplacian to functions that satisfy the Dirichlet
condition at the boundary $V=0$,
\begin{align}
-\bigtriangleup_D\psi(V)=-\bigtriangleup\psi(V) \ \textrm{ for } \
\psi(0)=0,
\end{align}
and then to close the operator $-\bigtriangleup_D$ in $L^2(\mathbb{R}_+,\dd
V)$. It can be shown that the generalized eigenfunctions are
\begin{align}
\psi_{\lambda}(V)=\ex^{i\sqrt{\lambda}V}-\ex^{-i\sqrt{\lambda}V}, \
\lambda\in \mathrm{Sp} (-\bigtriangleup_D)=\mathbb{R}_+,
\end{align}
and the propagator reads
\begin{align}
G_D(\eta,V,V')=\frac{\exp\left[-\frac{(V-V')^2}{4i\eta}\right]}{\sqrt{4\pi
i\eta}}-\frac{\exp\left[-\frac{(V+V')^2}{4i\eta}\right]}{\sqrt{4\pi i\eta}},
\label{propK0}
\end{align}
taking the original wave function from $0$ to $\eta$.

\subsection{Comparison with fast-time gauge}
\label{sec:p2}

\begin{figure}[t]
\includegraphics[width=0.46\textwidth]{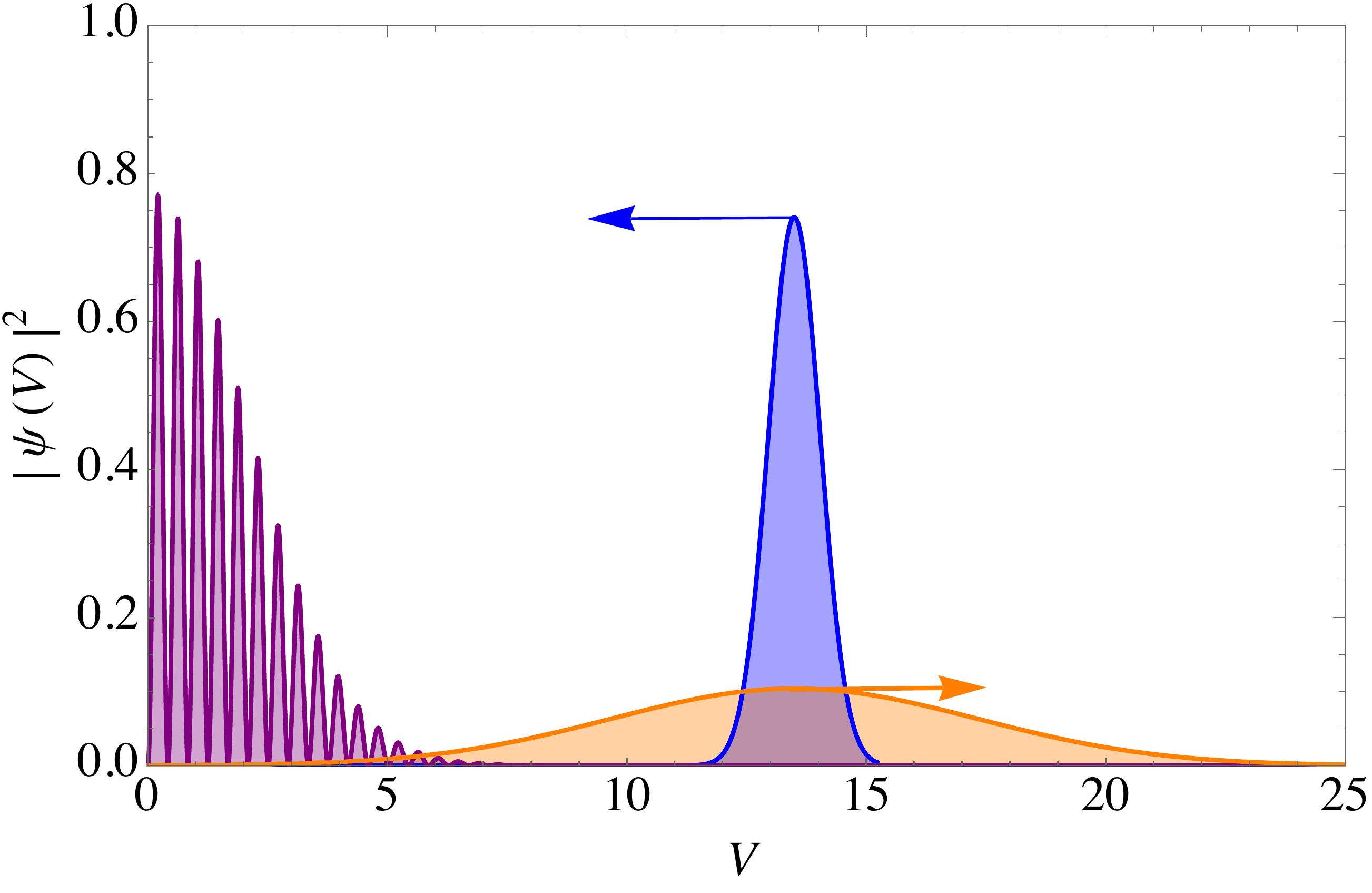}

\caption{The bouncing of a Gaussian wave packet against the endpoint
$V=0$ assuming the Dirichlet boundary condition. The wave packet momentum
is $k=7.5$ and times are as Fig.~\ref{FG}. The packet starts centered
around $V_0=15$ and variance $\sigma=1/2$ (arrow to the left); after one
unit of time it reaches the boundary where it interferes with itself and
after two units of time it returns to its initial position, though with
larger spreading  (arrow to the right).}

\label{bounce}
\end{figure}

Let us consider an initial wave function given by a Gaussian wave packet
centered at $V_0$ with standard deviation $\sigma$ and initial phase
$ikV$, namely
\begin{equation}
u_0(V) =
\frac{1}{\sqrt{\sqrt{2\pi}\sigma}}\exp\left[-\frac{(V-V_0)^2}{4\sigma^2} + ikV\right].
\end{equation}
In order for this waveform to satisfy the Dirichlet boundary condition at
$V=0$ and thus be an acceptable initial wave function, we consider its
odd part, i.e.,
\begin{equation}
\psi_0(V) = \frac{u_0(V) - u_0(-V)}{\mathcal{N}},
\end{equation}
where the normalization is given by 
\begin{equation}
\mathcal{N} = \sqrt{1 - \exp\left( - \frac{V_0^2+4k^2\sigma^4}
{2\sigma^2}\right) }.
\label{Norm1}
\end{equation}
Applying the propagator to this wave function gives us
\begin{equation}
\psi(V, \eta) =-
\frac{2 \exp\left({-ik^2\eta-\frac{V^2+V^2_\eta}{4\sigma\sigma_\eta}}\right)}
{\mathcal{N}(2\pi)^{1/4}\sqrt{\sigma_\eta}} 
\sinh\left(\frac{VV_\eta}{2\sigma\sigma_\eta}+ikV\right),
\label{sinh}
\end{equation}
where $V_\eta = V_0 + 2k\eta$ and $\sigma_\eta = \sigma+i\eta/\sigma$.
Rewriting the $\sinh$ above in terms of exponentials, it is easy to see
that we can complete the square in each exponent resulting in the
following expression
\begin{equation}
\begin{split}
\psi(V,\eta) &= \frac{\exp\left[+ik(V-k\eta)-\displaystyle\frac{\left(V -
 V_\eta\right)^2}{4\sigma\sigma_\eta}\right]}{\mathcal{N}(2\pi)^{1/4}
 \sqrt{\sigma_\eta}} \\
&\ \ \ -  \frac{\exp\left[-ik(V+k\eta)-\displaystyle\frac{\left(V +
 V_\eta\right)^2}{4\sigma\sigma_\eta}\right]}{\mathcal{N}(2\pi)^{1/4}
 \sqrt{\sigma_\eta}}.
\end{split}
\label{soltot}
\end{equation}
The above wave function \eqref{soltot} solves the Schr\"odinger equation
corresponding to a freely moving particle on the half-line with the
Hamiltonian $-\bigtriangleup_D$ with respect to our time variable $\eta$,
namely
\begin{align}\label{psi0K0}
i\frac{\partial}{\partial\eta}\psi=-\bigtriangleup_D\psi.
\end{align}
Disregarding the different phases, we end up with a linear combination of
two Gaussian wave packets centered on $V_\pm (\eta) = \pm(V_0+2k\eta)$
with spreading variance $\sigma(\eta) = \sqrt{\sigma^2 +
\eta^2/\sigma^2}$. Its evolution is shown in Fig. \ref{bounce}. Contrary
to the fast-gauge time, the boundary is now reached by the wave packet
within a finite time interval, which must bounce in order to preserve the
unitarity.

The quantum model presented here with $\hat{H}=-\Delta$ is based on an
implementation of canonical quantization rules in the case of the
half-line. This approach, however, is not fully satisfactory as it
assumes the momentum on the half-line, $(-i\partial_V)$, to be one
of the basic operators despite the fact that it is not a self-adjoint
operator. As a related problem, the quantum Hamiltonian, $-\Delta$, is
not an essentially self-adjoint operator either; therefore its domain is
confined to a certain dense subspace of the full Hilbert space of the
model and its action beyond this restricted domain is redefined in order
to make it self-adjoint. This procedure is highly ambiguous and produces
a significant technical inconvenience: once the action of  $\hat{H}$ is
redefined, its commutation with other operators can no longer be
determined from its representation as a differential operator, i.e. as
$-\Delta$. This is a drawback because, as we show below, by making
use of commutation rules, one is able to prove the existence of a symmetry
in the quantum dynamics of the Bianchi I model, which enables one to
immediately obtain the evolution of some operators.

Therefore, in what follows, we implement {\it affine quantization} in
which the nonself-adjoint momentum operator is replaced with the
self-adjoint dilation operator,\footnote{The name ``affine" is due to the
fact that the dilation and position operators generate the unitary
irreducible representation of the affine group of the real line.}
$D\equiv \frac12 \left( V \frac{1}{i}\partial_V +
\frac{1}{i}\partial_VV\right)$. The dilation and position operators
provide two basic operators from which any compound operator such as the
Hamiltonian can be obtained. In this case one is faced with the ordering
issue. Nevertheless, all the orderings are shown to produce the same form
of the quantum Hamiltonian and a wide class of them are self-adjoint
operators which produce a unique dynamics and can be represented as
differential operators.

\subsection{Affine quantization}\label{quantize_sec}

Our Hamiltonian thus reduces to $\pi_V^2$, which can be
classically expressed in terms of the symmetric combination $D\equiv \frac12
\left( V \pi_V + \pi_V V\right)$ as $\pi_V^2 \sim D^2/ V^2$. Upon quantization,
it is well known that this leads to an ambiguity as the order of the
corresponding operators becomes relevant.

Indeed, with the canonical commutation relation $[ \hat
V,\hat\pi_V] = i$, one finds $[ \hat V,\hat D] = i\hat
V$, so that one can express the Hamiltonian in the symmetric form
\begin{equation}
\hat H = \hat V^\alpha \hat D \hat V^\beta D \hat V^\alpha \quad
\hbox{with} \quad 2\alpha + \beta =-2,
\label{Halpha}
\end{equation}
for generic values of $\alpha$. Using the commutation relation
$$
[ \hat V^\alpha,\hat D] = i\alpha\hat V^\alpha
$$
to move all the $\hat V$ factors to the left, and going back to
$\hat\pi_V$ in the final result leads to
\begin{equation}
\hat H = \hat\pi_V^2 + \frac{\Ka(\alpha)}{\hat V^{2}}
\quad\hbox{with}\quad \Ka(\alpha) \equiv \alpha^2 + 2 \alpha + \frac34.
\label{HKalpha}
\end{equation}
It turns out this new Hamiltonian is essentially self-adjoint if
$\Ka(\alpha) > \frac34$ (see Ref.~\cite{Reed:1975uy}, page 161), i.e. for
$\alpha >0$ or $\alpha<-2$. We assume in what follows that $\alpha$ is
chosen to ensure the required self-adjointness of the
Hamiltonian.\footnote{One recovers exactly the same result by assuming
the correspondence $\pi_V^2 \mapsto \hat V^s \hat\pi_V \hat V^{-2s}
\hat\pi_V \hat V^s$, leading to a similar potential term: $\pi_V^2 \mapsto
\hat\pi_V^2 +s \hat V^{-2}$, and a self-adjoint Hamiltonian provided
$s>3/4$.} The appendix shows that the wave packet behavior
in this case is essentially the same as that illustrated on Fig.~\ref{bounce},
Eq.~\eqref{psiVeta} showing the generalization for $K(\alpha)\not=0$ of
Eq.~\eqref{soltot}.

Let us note at this point that one could expect self-adjointness to be
naturally derived from some other physically justified assumptions and
not, as we are here proposing, imposed as a mathematical input. One could
however argue in the opposite direction: consider for instance the simple
case of the Hydrogen atom. Applying the correspondence principle to the
classical Hamiltonian already yields a self-adjoint operator whose
quantization permits one to calculate the energy levels and compare those
with data. Expanding the electron momentum operator into radial and
angular components, one could apply our ordering procedure to find an
extra potential of the form $C/r^2$, the arbitrary constant $C$ being
then fixed by comparison of the resulting (different) energy levels with
the data. One could then argue that the operator ordering choice can be
determined experimentally, even in situations where the original
Hamiltonian is already self-adjoint.

\subsection{General time evolution}\label{integrate_sec}

Let us begin by working in the Heisenberg representation and discuss time
evolution of the relevant operators. Using the usual relation
$[\hat\pi_V,f(\hat V)] = -i\displaystyle\frac{\dd f(\hat V)}{\dd \hat V}$
together with the commutation relations $[\hat\pi_V,\hat H] = 2iK \hat
V^{-3}$ and $[\hat V,\hat H] = 2i\hat \pi_V$, one readily finds that the
algebra made with $\hat V^2$, $\hat D $ and $\hat H$ is closed, namely
\begin{equation}
\left\{ \begin{array}{rcl}
[ \hat V^2,\hat H ] & = & 4 i \hat D, \cr
[ \hat D,\hat H ] & = & 2 i \hat H, \cr
[ \hat V^2,\hat D ] & = & 2 i \hat V^2,
\end{array} \right.
\label{algebra}
\end{equation}
leading to the Heisenberg equations of motion for the time
development of the operators, namely
\begin{equation}
\frac{\dd}{\dd\eta}\hat V^2 = -i [\hat V^2, \hat H ] = 4 \hat D,
\label{Q2dot}
\end{equation}
for the squared volume operator, and
\begin{equation}
\frac{\dd}{\dd\eta}\hat D = -i [\hat D, \hat H ] = 2 \hat H,
\label{Ddot}
\end{equation}
with $\hat H$ being a constant operator.

\begin{figure}[t]
\includegraphics[width=0.46\textwidth]{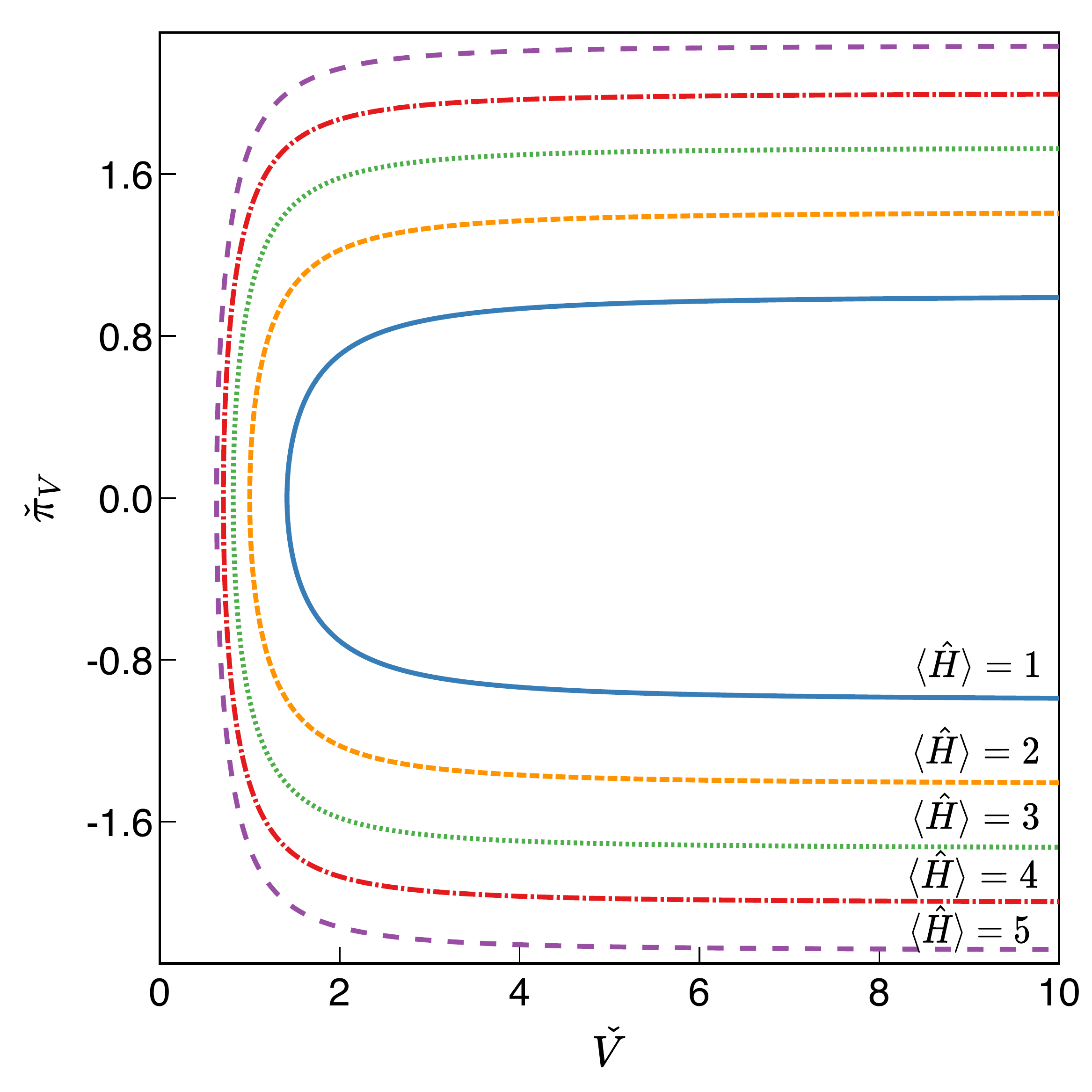}

\caption{Effective phase space trajectories \eqref{vtpt} for the
effective Hamiltonian \eqref{Heff}, taking $\langle\hat H\rangle \in
[1,5]$ from the inside to the outside of the graph, as indicated, full
($\langle\hat H\rangle =1$), dashed ($\langle\hat H\rangle =2$), dotted
($\langle\hat H\rangle =3$), dot-dashed ($\langle\hat H\rangle =4$) and
long-dashed ($\langle\hat H\rangle =5$). We use $\Ka(\alpha)=2$ for the
plot.}

\label{vpt00}
\end{figure}

Because of the constancy of $\hat H$ in time, one can explicitly
integrate \eqref{Ddot}, namely
\begin{equation}
\hat D(\eta) = 2 \hat H \eta + \hat D(0),
\label{Dsol}
\end{equation}
which, once plugged into \eqref{Q2dot}, leads to
\begin{equation}
\hat V^2 = 4 \hat{H} \eta^2 + 4 \hat D(0) \eta + \hat V^2(0).
\label{Q2sol}
\end{equation}
The expectation values of these operators follow simple trajectories,
whatever the state one integrates over. They read
\begin{equation}
\langle \hat D (\eta)\rangle = 2 \langle \hat H\rangle \eta + d_0
\label{AverageD}
\end{equation}
where we have set $d_0 \equiv \langle \hat D (0)\rangle$, and
\begin{equation}
\langle \hat V^2 (\eta)\rangle = 4 \langle \hat H\rangle \eta^2 + 4 d_0
+ v_0^2, \label{AverageQ2}
\end{equation}
with $v_0^2 \equiv \langle \hat V^2 (0)\rangle$.

Shifting the time variable to $t=\eta + d_0/(2\langle \hat H\rangle)$
and setting $V_0^2 = v_0^2 -d_0^2/\langle \hat H\rangle$ (assuming
$v_0^2\langle \hat H\rangle \geq d_0^2$), one can now define
semiclassical variables $\check{V}(t)$ and $\check{\pi}_V(t)$ through
\begin{equation}
\check{V}(t) = \sqrt{\langle \hat V^2 (t)\rangle} \quad\hbox{and}\quad
\check{\pi}_V(t) = \frac{\langle \hat D (t)\rangle}{\check{V}(t)},
\label{vpt}
\end{equation}
and finally obtain a set of trajectories in phase space labeled by the
arbitrary time $t$, namely
\begin{equation}
\begin{array}{rcl}
\check{V}(t) &=&\sqrt{4\langle \hat H\rangle t^2 + V_0^2},\cr
\check{\pi}_V(t)&=&\displaystyle\frac{2\langle \hat H\rangle t}{\sqrt{4\langle \hat
H\rangle t^2 + V_0^2}}.
\end{array}
\label{vtpt}
\end{equation}
Each trajectory is thus labeled by two parameters, namely the average
value of the Hamiltonian $\langle \hat H \rangle$ and the minimum
volume $V_0$, and indeed, we have
\begin{equation}
\langle \hat H \rangle = \frac{\check{V}^2(t) \check{\pi}_V^2(t)}{\check{V}^2(t)-V_0^2} = \check{\pi}_V^2(t) +
\frac{\Ka}{\check{V}^2(t)},
\label{Heff}
\end{equation}
provided one sets $\Ka = \langle \hat H \rangle V_0^2$. 

\begin{figure}[t]
\includegraphics[width=0.46\textwidth]{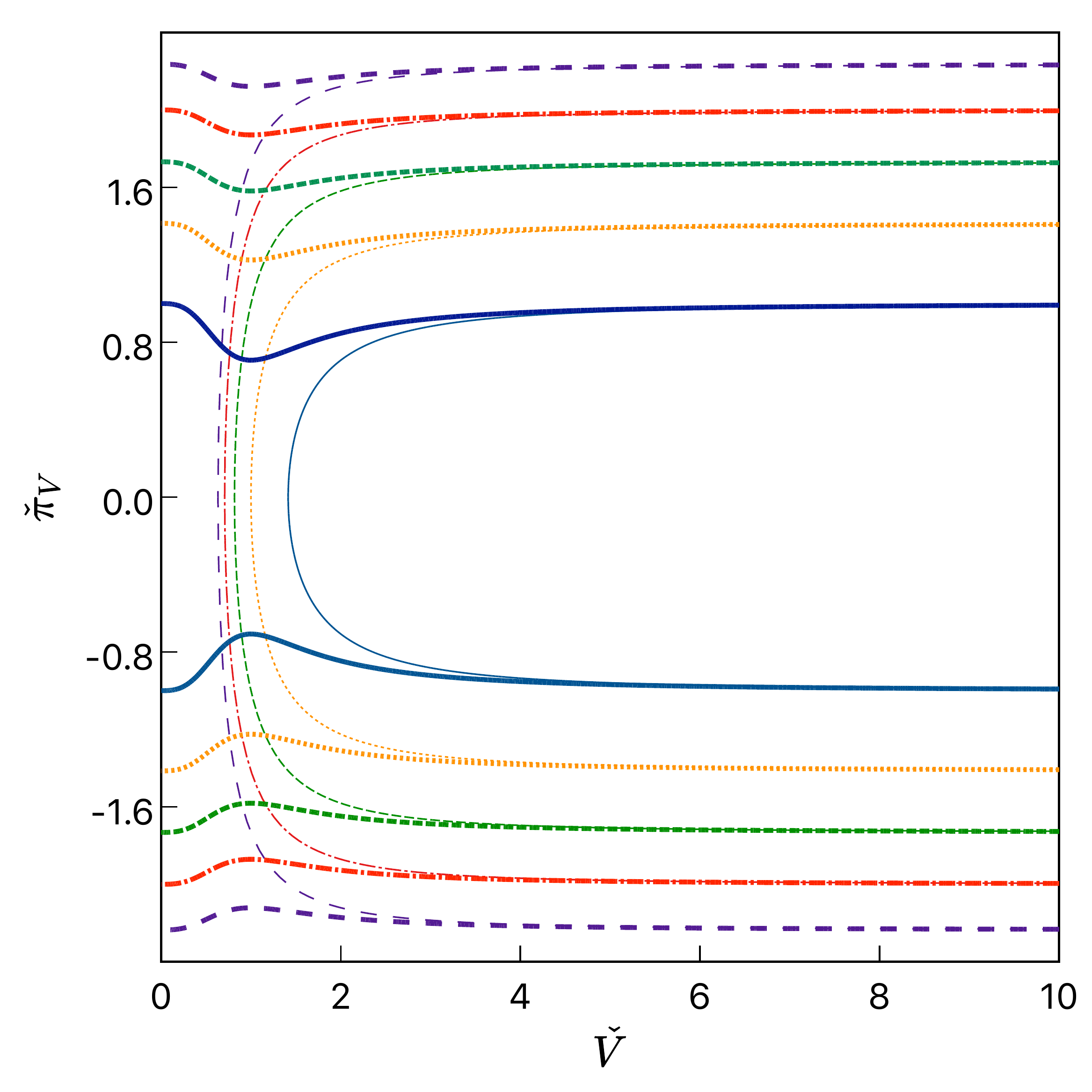}

\caption{Same as Fig. \ref{vpt00} after application of the delay function
\eqref{FastSlowTT}. The originally regular trajectories (thin lines) are
now all singular (thick lines).}

\label{vptSing}
\end{figure}

It is interesting to realize that the phase portrait for the regular case
of the slow-gauge time is transformed, using the delay function
\eqref{FastSlowTT}, into singular solutions: as shown on
Fig.~\ref{vptSing}, all solutions now either terminate in or originate
from a singularity $V\to 0$.

\begin{figure*}[t]
\includegraphics[width=0.46\textwidth]{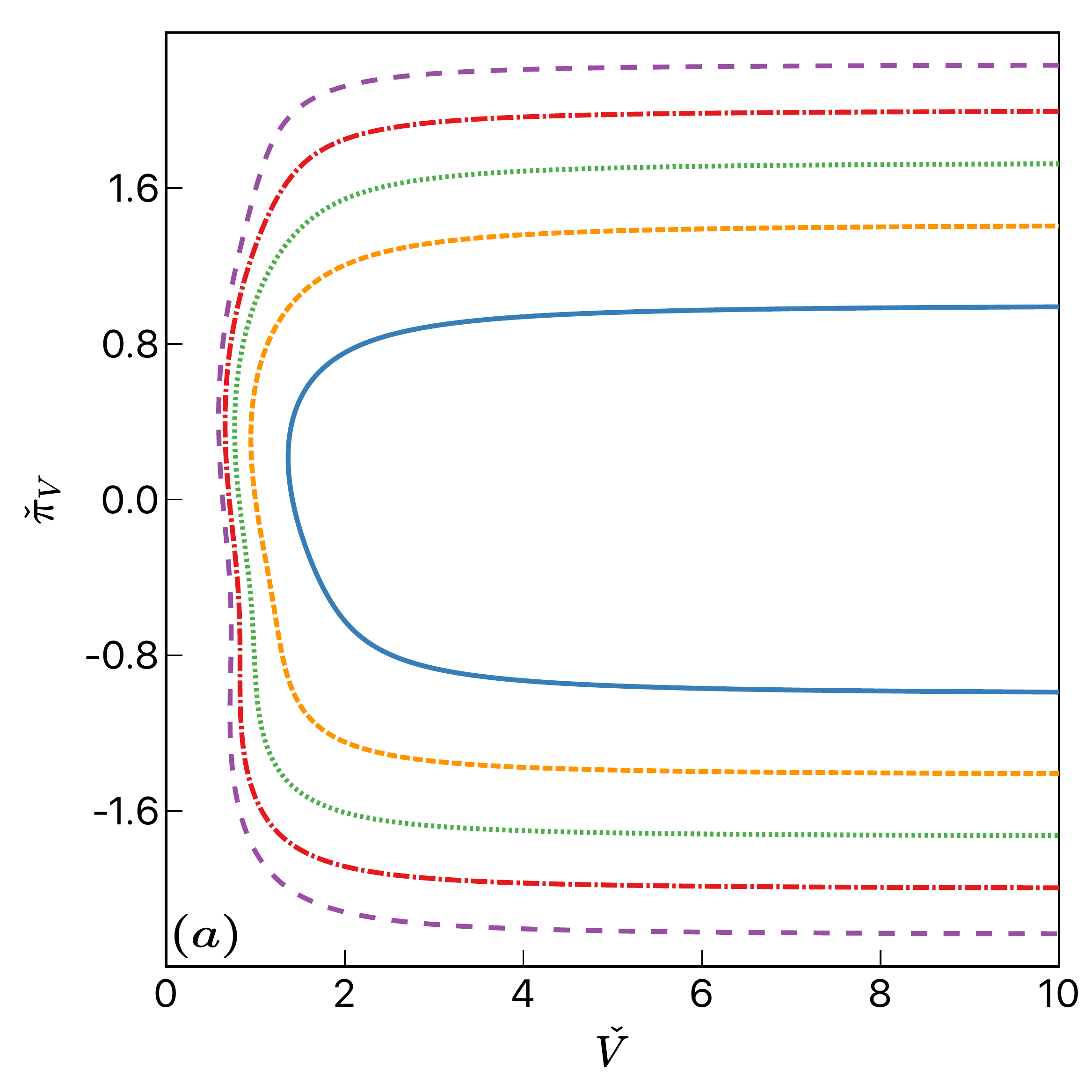}
\includegraphics[width=0.46\textwidth]{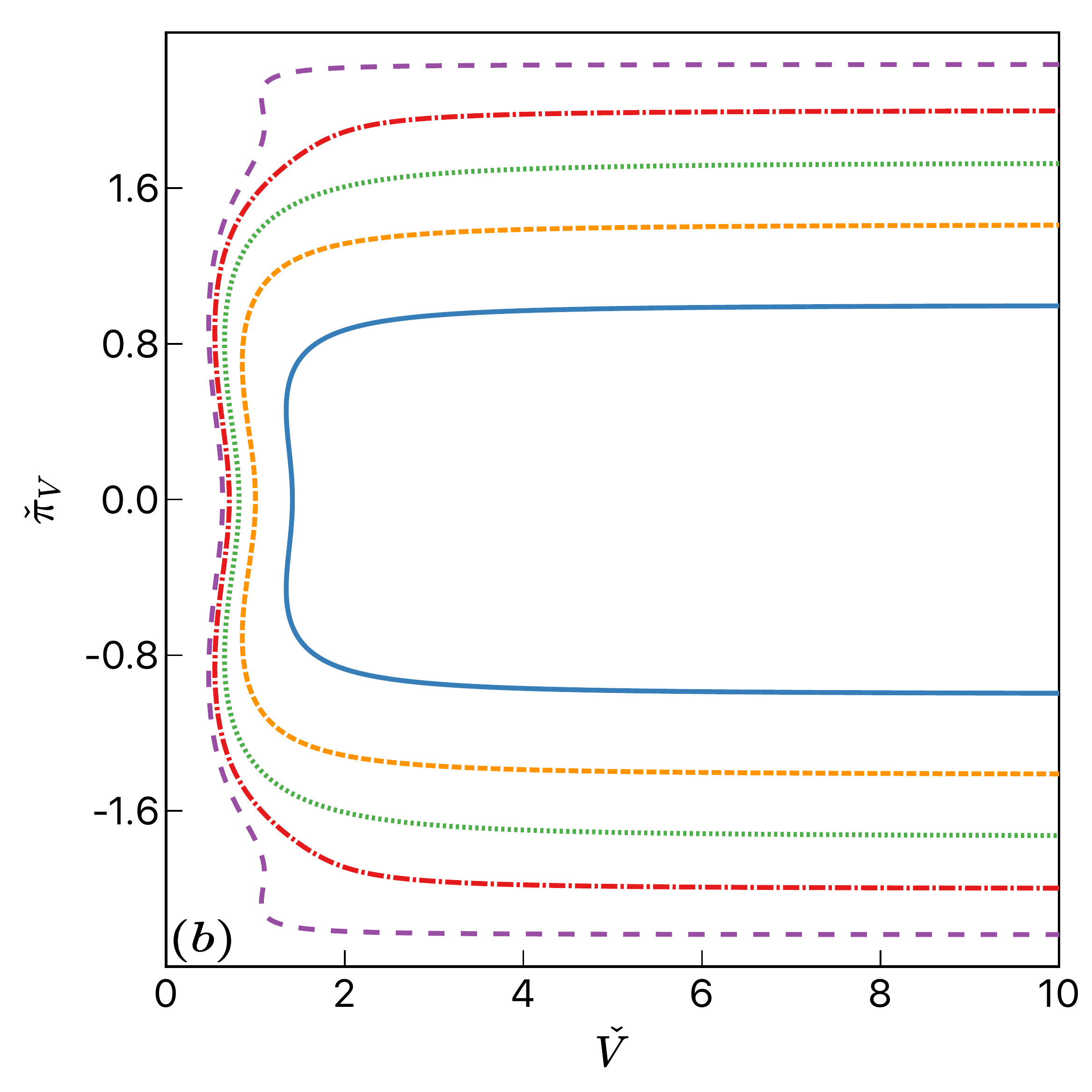}
\includegraphics[width=0.46\textwidth]{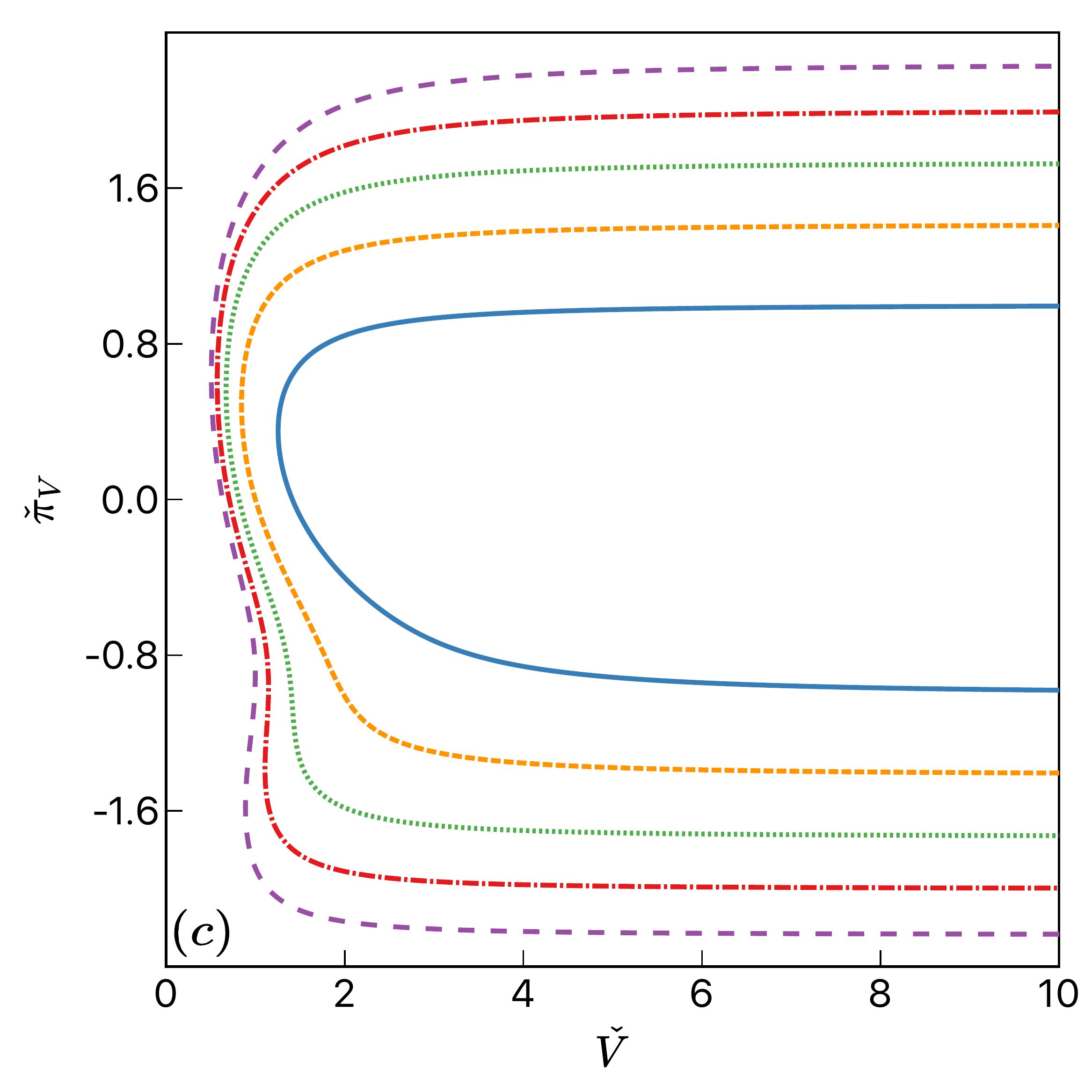}
\includegraphics[width=0.46\textwidth]{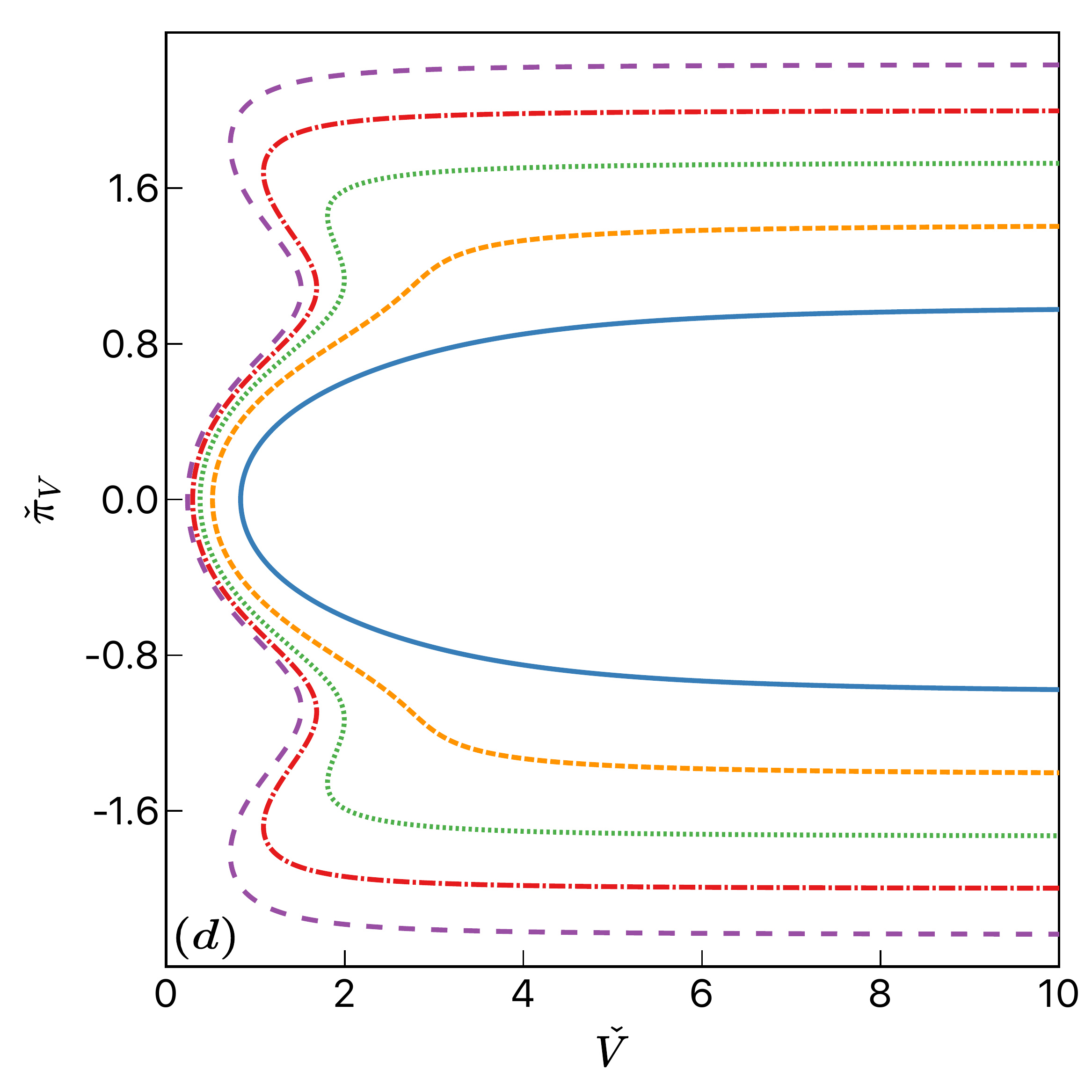}

\caption{Same as Fig. \ref{vpt00} after application of the delay
functions $\Delta = V \ex^{-2|\pi_V|/3} \sin (3 V \pi_V)/(10 \pi_V)$
$(a)$, $\Delta = V(\pi_V - 10^{-0.2} \pi_V^3 + \pi_V^5/10)$ $(b)$,
$\Delta = 10^{-0.5} V \sin (2 \pi_V)/\pi_V$ $(c)$ and $\Delta = 10^{-0.5}
(V+1) \cos (3 \pi_V)/\pi_V$ $(d)$. It can be checked that these delay functions
satisfy the requirement \eqref{etaprimeOK}. The new trajectories happen
to be not necessarily symmetric like their counterpart of Fig.~\ref{vpt00}.}

\label{vpt02}
\end{figure*}

\subsection{Comparison of different slow-gauge dynamics}

So far we have shown quantization of the model in a single internal time,
$\eta$. As we have shown in Sec. \ref{clock_sec}, all other choices of internal time,
denoted by $\eta'$, can lead to formally the same Hamiltonian framework
provided that a suitable choice of the new canonical pair, $V'$ and
$\pi_V'$, defined in Eq. (\ref{newcan}), is made. In this case, the
quantization introduced in Sec. \ref{quantize_sec} and the subsequent integration of the quantum motion given in Sec. \ref{integrate_sec} can be repeated simply by replacing the
labels of the canonical variables, $V\rightarrow V'$ and
$\pi_V\rightarrow \pi_V'$. Actually, there is a much better reason than
mere technical convenience for repeating the quantization in this
particular manner: since the Hamiltonian frameworks are formally
identical, the constants of motion derived within them must be formally
identical functions of the respective basic variables and internal time. Hence,
repeating the quantization in all internal times promotes the
constants of motion to the same operators irrespectively of the choice of
internal time. On the other hand, the constants of motion enjoy a
physical interpretation that must not depend on the particular choice of
time. Therefore, the quantization of the system is in this sense unique
for all internal frames. One also notices that since the number of elementary constants of
motion is equal to the dimensionality of the phase space, the quantization
cannot be {\it more} unique, i.e. it is completely determined by the
quantization of the constants. 

We expect that, contrary to the case of constants of motion, quantization
of dynamical observables in general leads to different operators for
different internal times. This is the reflection of the fact already
mentioned in the introduction that dynamical observables are not
gauge invariant in Hamiltonian constraint systems. For a more detailed
discussion of these and related issues, we refer the reader to
\cite{Malkiewicz:2017cuw}.

Let us explain our approach to making the comparison between quantum
dynamics in different internal times. First, we note that all quantum
dynamics are placed in a single Hilbert space that carries a unique
quantum representation of constants of motion. Second, the quantum
dynamics viewed as a curve in the Hilbert space is actually unique
because the quantum Hamiltonian generating the dynamics is a quantum
constant of motion that is unique in all internal times. Third, to
describe the quantum dynamics, one needs operators that do not commute
with the Hamiltonian and are not quantum constants of motion. However,
such operators are exactly the operators which correspond to different
physical observables in different internal clocks. Therefore, using the
same operator(s) for the purpose of describing the time evolution of the
quantum system must be complemented by a physical interpretation of the
operator(s), which must depend on the choice of internal time. Hence,
formally the same dynamics in the Hilbert space will render different
physical portraits for different internal times. The extent to which the
physical portraits differ is the result of the choice of internal time
and we refer to it as time effect.

Finally, let us notice that one could instead choose the same physical
observable and determine the respective operators in each internal time
and then compare the dynamics of these operators. Such an approach, in
principle valid, is technically much more involved or even impossible to
apply if a given physical observable does not enjoy a self-adjoint
representation in a given internal time.

Let us now establish a concrete computational scheme for the comparison
method outlined above. Equation \eqref{vtpt} defines the semiclassical
portrait of the dynamics of the model in terms of $\check{V}$ and
$\check{\pi}_V$ in one internal time. As discussed above, the
quantization of the same model in another internal time yields the same
form of the semiclassical portrait except for that now the coordinates
are  $\check{V}'$ and $\check{\pi}_V'$ rather than $\check{V}$ and
$\check{\pi}_V$. In order to compare the two portraits we use the
relation between the basic observables given in Eq. \eqref{newcan}, i.e.
\begin{align}
\check{\pi}_V=\check{\pi}_V',~~\check{V}=\check{V}'+\check{\pi}_V'
\Delta(\check{V}',\check{\pi}_V').
\end{align}
By choosing various delay functions $\Delta(\check{V}',\check{\pi}_V')$
we are able to generate infinitely many new semiclassical portraits all
of which describe the quantum dynamics of the Bianchi I model in terms of
the same observables $\check{V}$ and $\check{\pi}_V$ but produced with
different internal times.

Figure \ref{vpt02} shows various cases for which we have picked arbitrary
but acceptable delay functions $\Delta_i(\check{V}',\check{\pi}_V')$. It
is clear from these graphs that the ``actual'' motion in phase space can
be for the most part arbitrary. In particular, it is neither necessarily
symmetric in the $(\check{V}',\check{\pi}_V')$ plane. Moreover, one can
even find a minimum volume at points for which the momentum is
nonvanishing, thereby ruining the usual interpretation of the latter as
the Hubble factor

\begin{figure}[t]
\includegraphics[width=0.46\textwidth]{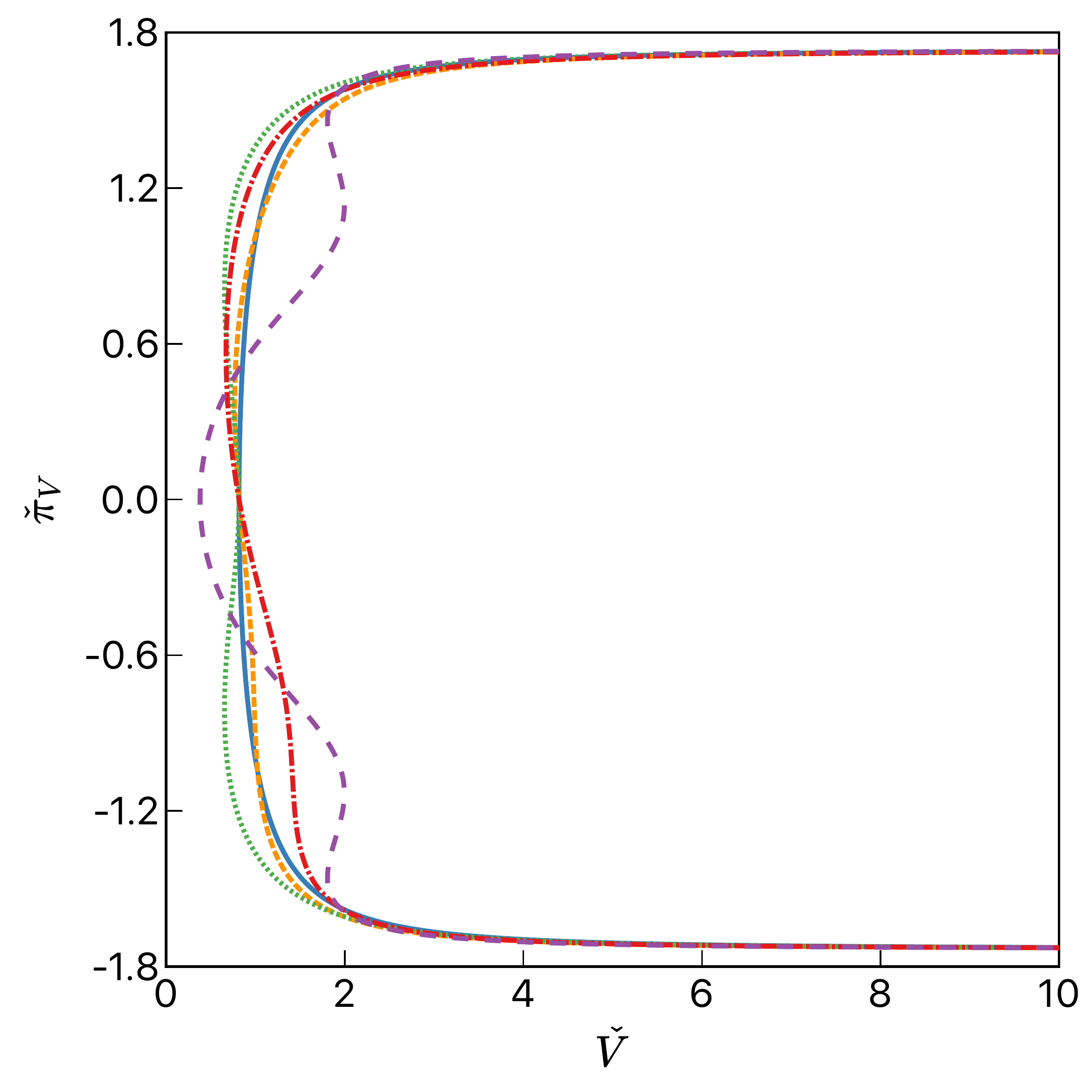}

\caption{Comparison between five delay functions $\Delta$ (shown on Fig.
\ref{vpt02}) and the original time phase space trajectories from Fig.
\ref{vpt00}, illustrated here on the case $\langle \hat H \rangle= 3$.
The asymptotic behaviors being identical, the classical limit appears to
be well defined whatever the delay function used, which is significant
only around the bouncing epoch.}

\label{vpt05}
\end{figure}

\section{Conclusions}

We studied the empty Bianchi I universe to exemplify the use of a clock
in quantum cosmology. Solving the classical Hamilton equations, we find
two different categories of clocks, dubbed fast and slow-gauge times. The
fast-gauge time appears in a more natural way in the canonical one-form,
and yields a singular classical motion, although it requires an infinite
amount of fast-gauge time to reach it (hence the gauge name). It has been
conjectured, and we provide an explicit example, that canonical
quantization cannot remove the singularity, the wave function eventually
evolving toward a Dirac distribution at vanishing volume.

Solving the constraint using a more sophisticated solution provides
another category of clocks, dubbed slow-gauge times. Classically, such
clocks are slow in the sense that the singularity is now reached in a
finite amount of time. The question of time is now manifested by the fact
that there exist many choices, all involving a delay function thanks to
which new sets of canonical variables may be defined.

The main difference between fast and slow-gauge times, in the Bianchi I
case, resides in the domain of definition of the variables. In the fast
case, the evolution is naturally unbounded, the Hamiltonian being that of
a free particle on the full real line, whereas the slow-gauge time yields
a similar evolution but only on the half-line. Up to some technical
points regarding the self-adjointness of operators, this permits one to
resolve the classical singularity through quantum mechanical effects.

We show that in the Heisenberg picture, it is possible to explicitly
solve the relevant operators (Hamiltonian, dilation and square of the
volume) as functions of time, allowing to draw phase portraits. We then
find that, in a way mostly independent of the explicit choice of state
itself (which is an advantage of the Heisenberg picture over
Schr\"odinger's), the phase space trajectories are always similar,
depending on the eigenvalue of the Hamiltonian. A Gaussian wave packet
evolution shows exactly the same behavior, as expected.

Shifting to different times by picking arbitrary delay functions, one
finds that the phase space trajectories depend strongly on the time
choice only when quantum effects are relevant, i.e. close to the bouncing
point (minimum of the volume). However, we also show that there exists an
asymptotic regime in which the semiclassical motion is a good
approximation and which does not depend on the choice of time
(Fig.~\ref{vpt05}). These results are in agreement with earlier results
on the time issue for the Friedmann model filled with radiation
\cite{Malkiewicz:2015fqa}. It could thus be conjectured that the question
of time in a quantum cosmological setting is naturally resolved in the
classical domain provided such a regime exists. In other words, time
would cease to be a relevant physical object in the quantum gravitational
realm, recovering its meaning only for configurations for which the use
of general relativity is appropriate. At the moment, one needs to
implement a time parameter to order events, but it may not be necessary
in a more complete theory.

Since the empty Bianchi I model with internal time effectively becomes
one dimensional, it is fully justified to ask whether the results
obtained in this paper are restricted to one-dimensional models and no
longer hold when an additional degree of freedom is present and the space
of solutions has more structure. This question was in fact to some extent
already investigated for the case of the Bianchi I with a fluid playing
the role of  the internal time variable \cite{Malkiewicz:2016hjr} and the
results obtained there appear to be in full agreement with the ones
presented herein. Nevertheless, it seems to us that more such studies are
desirable.

The next and related question that needs be asked concerns perturbations,
and in particular whether they also enjoy a unique classical limit
independent of the choice of time. If true, such a statement would permit
deriving ``matching conditions'' (as discussed, e.g., in
Ref.~\cite{Martin:2002ar}); we postpone a discussion of perturbations in
a vacuum Bianchi I universe for future work.

\acknowledgments

The project is cofinanced by the Polish National Agency for Academic
Exchange and PHC POLONIUM 2019 (Grant No. 42657QJ).

S.~D.~P.~V. acknowledges financial aid from the PNPD/CAPES (Programa
Nacional de P\'os-Doutorado/Capes, Grant No. 88887.311171/2018-00).

P.~P. thank the Labex Institut Lagrange de Paris (Grant No.
ANR-10-LABX-63), part of the Idex SUPER, within which this work was
partly done. P.~P. is hosted at Churchill College, Cambridge, where he is
partially supported by a fellowship funded by the Higher Education,
Research and Innovation Department of the French Embassy to the United
Kingdom. P.~P. also expresses a special thanks to the Mainz
Institute for Theoretical Physics (MITP) of the DFG Cluster of Excellence
PRIMSMA+ (Grant No. 39083149), for its hospitality and support.

\begin{figure}[t]
\includegraphics[width=0.46\textwidth]{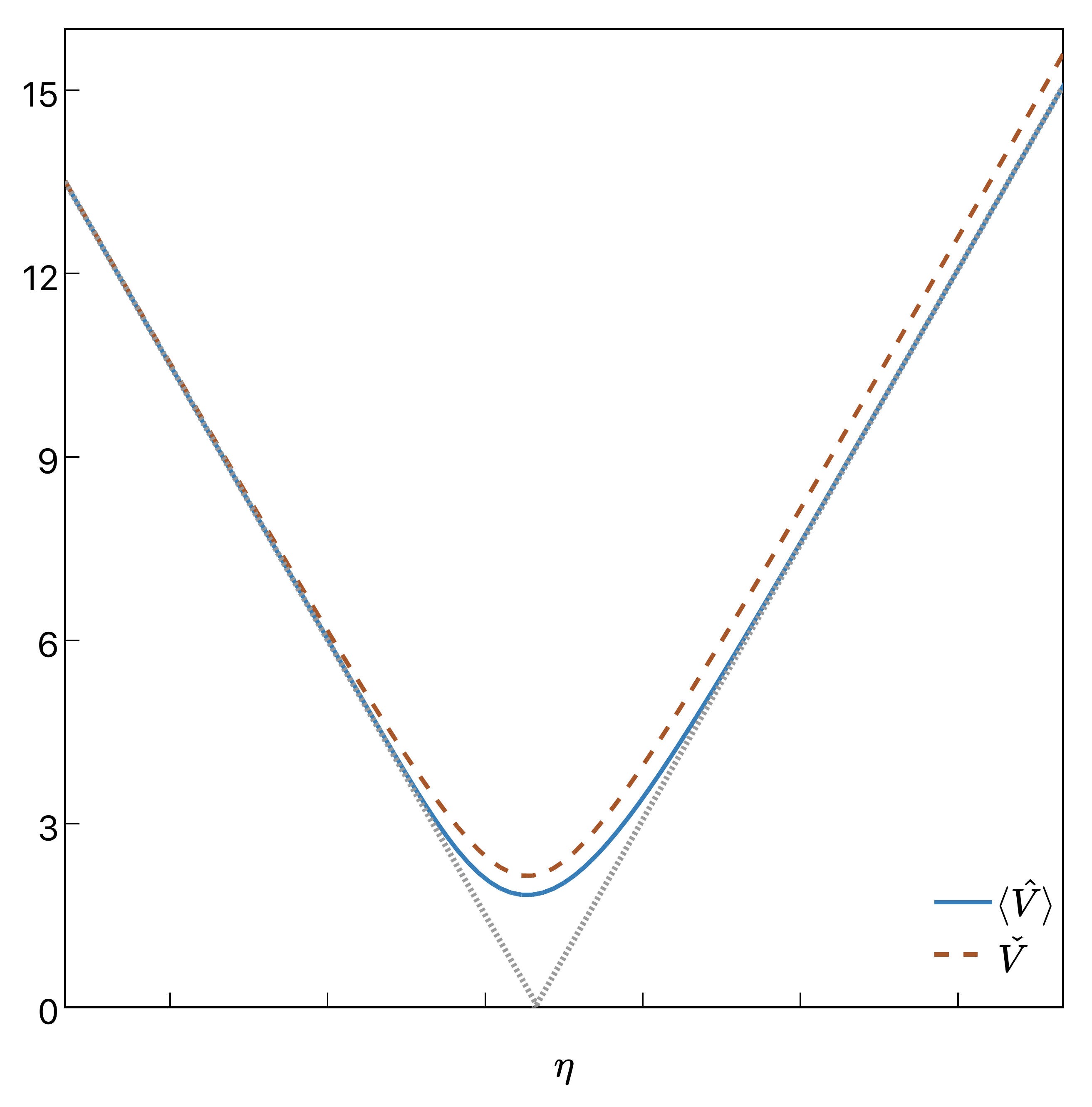}

\caption{Variation of the mean value $\langle \hat{V}\rangle$ with the
conformal time $\eta$ using the solution \eqref{psiVeta} (full line),
compared with the semiclassical approximation given by the $\check{V}$
\eqref{vtpt} (dashed line). Also shown is the classically singular
trajectory (dotted line).}

\label{V2eta}
\end{figure}

\appendix*

\section{Wave packet for the affine case}

In Sec.~\ref{sec:p2}, we presented the evolution of a Gaussian wave
packet for the simple case where the Hamiltonian is given by $\hat{H} =
-\bigtriangleup_D$. In this appendix, we discuss the equivalent situation
for the affine case \eqref{Halpha} for which we obtained the Hamiltonian
\eqref{HKalpha}. Still using the representation for which $\hat{V}$ is
multiplicative, i.e.,
$$
\hat{V}\psi(V) = V\psi(V), \qquad \hat\pi_V \psi(V) = -i\partial_V\psi(V),
$$
the Hamiltonian operator reads,
\begin{equation}
\hat H = -\partial_V^2 + \frac{\Ka (\alpha)}{V^2}.
\end{equation}
All the eigenfunctions of this operator satisfy the Dirichlet
conditions $\psi_\ell(0) = 0$, and they read
\begin{equation}
\begin{split}
\psi_\ell &= \sqrt{\ell V}J_\nu (\ell V),  \qquad \ell \in \mathrm{Sp}
\left(\hat H\right)=\mathbb{R}_+, \\
\nu &= \frac{\sqrt{1+4K(\alpha)}}{2},\qquad \hat H\psi_\ell =\ell^2 \psi_\ell ,
\end{split}
\end{equation}
where $J_\nu(x)$ is the Bessel function of the first kind.

\begin{figure}[t]
\includegraphics[width=0.46\textwidth]{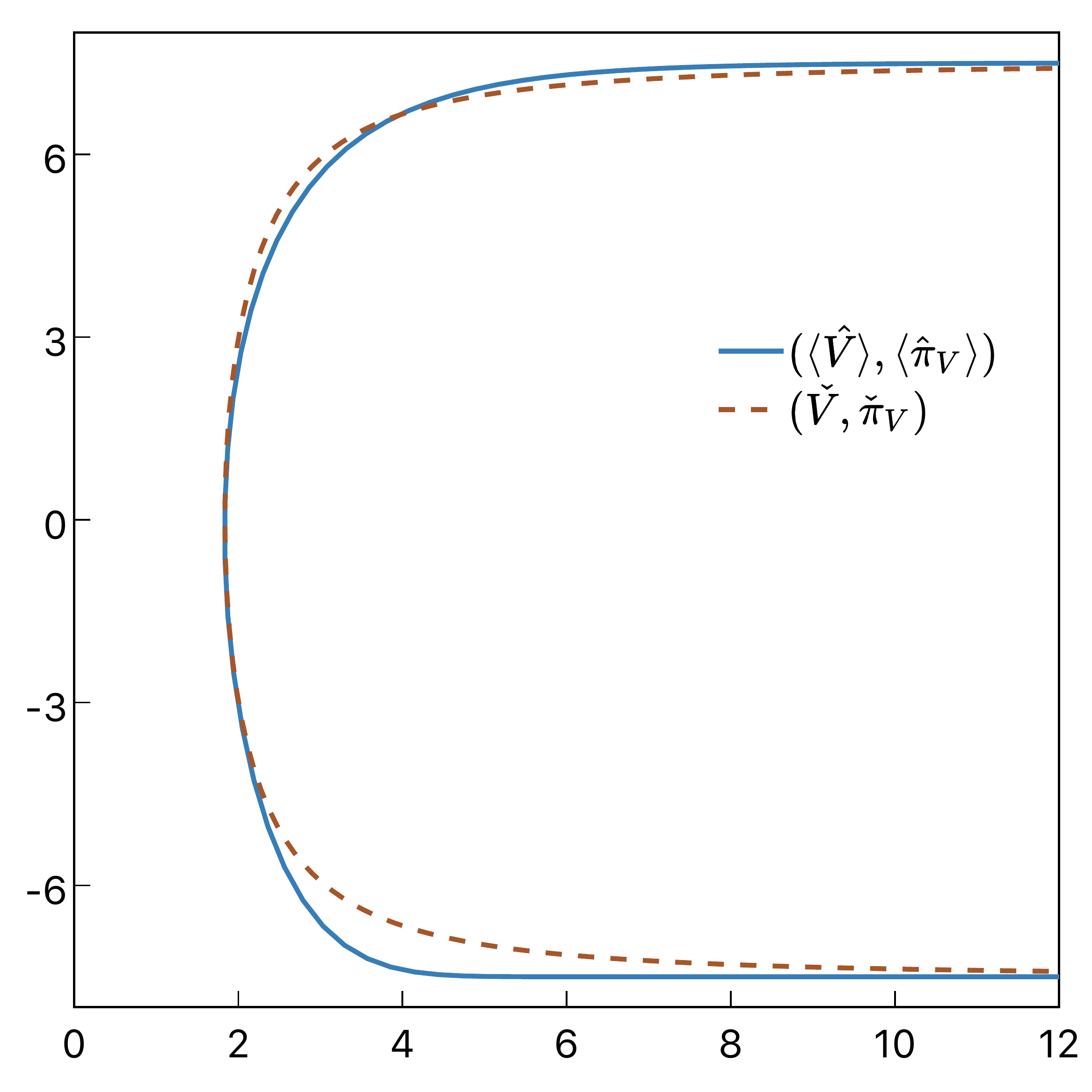}

\caption{Phase space evolution comparing the $\left(\langle
\hat{V}\rangle, \langle \hat{\pi}_V \rangle \right)$ trajectory using
\eqref{psiVeta} (full line), with the symmetric semiclassical
approximation given by the $\left(\check{V}, \check{\pi}_V \right)$
\eqref{vtpt} (dashed line). The asymptotic behavior is identical.}

\label{pvpv}
\end{figure}

The propagator $G$ is given by the integral of the eigenfunctions over the
spectrum, namely
\begin{equation}
G(\eta, V, V^\prime) = \sqrt{VV^\prime}\int_0^\infty\mathrm{d}\ell\; 
\ell \ex^{-i\ell^2\tilde\eta}J_\nu(\ell V)J_\nu(\ell V^\prime),
\end{equation}
where $\tilde{\eta} = \eta(1\mp i\epsilon)$ for $\epsilon > 0$ gives the
correct propagator prescription after taking $\epsilon\to0$ for $\eta >
0$ and $\eta < 0$ respectively. This integral can be done analytically
using Weber second integrals
(see~\cite[Sec.~10.22.67]{Olver:2010:NHM:1830479}),
\begin{equation}
G(\eta, V, V^\prime) = \frac{\sqrt{VV^\prime}}{2i\tilde{\eta}}
\exp\left(-\frac{V^2+V^{\prime2}}{4i\tilde{\eta}}\right)
I_\nu\left(\frac{VV^\prime}{2i\tilde{\eta}}\right),
\label{propK}
\end{equation}
where $I_\nu(x)$ is the modified Bessel function of the first kind. Note
that $\Ka (\alpha) \to 0 \Rightarrow \nu \to 1/2$, and the Bessel function
reduces to the $\sinh$: $I_{1/2}(x) = \sinh(x)/\sqrt{\pi x / 2}$.
Substituting this expression in the propagator above then reproduces, up
to an irrelevant phase, Eq.~\eqref{propK0}.

\begin{figure}[t]
\includegraphics[width=0.46\textwidth]{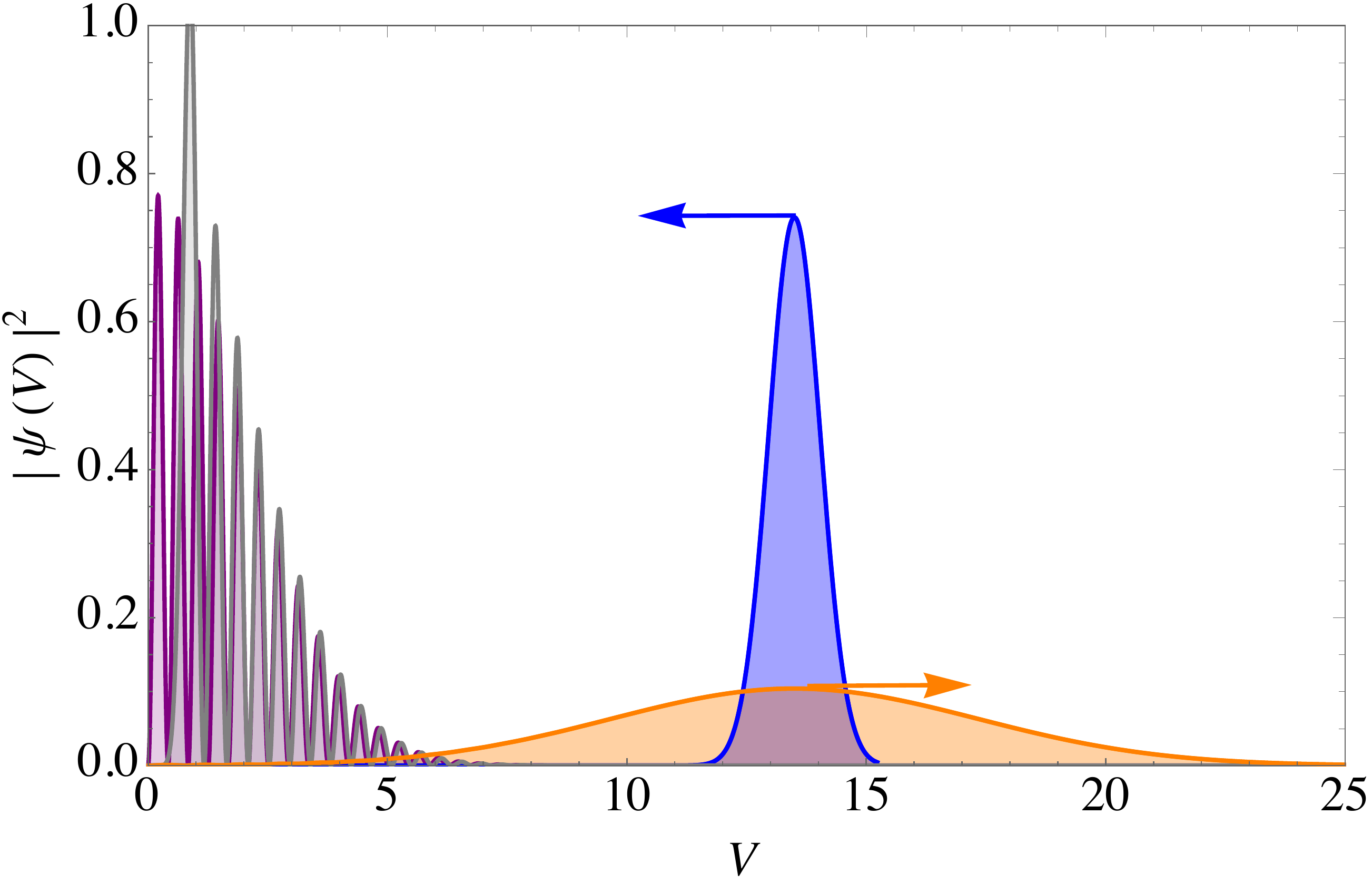}

\caption{Same as Fig.~\ref{bounce} for the affine case with $\Ka=7/4$,
i.e. the solution \eqref{psiVeta}. Two sets of oscillations are shown,
namely those stemming from the free case $\nu=\frac12$ discussed in
Sec.~\ref{sec:p2}, and the new situation with $\nu=2$. Although there are
important differences near the bounce, the asymptotic behavior does not
depend on the operator ordering choice, as expected for a meaningful
semiclassical approximation; this comes from the fact that the potential
becomes negligibly small far from the bounce.}

\label{bounce2}
\end{figure}

One useful property of the propagator~\eqref{propK0} is that its integral
over a Gaussian  distribution generates another Gaussian distribution, as
we have seen in Sec.~\ref{sec:p2}. Although the integral of the
propagator~\eqref{propK} does not have the same property when integrated
over Gaussian distribution, it is still possible to choose a different
initial wave packet that reduces to our earlier choice~\eqref{psi0K0} when
$\nu\to1/2	$ and retains the same functional form when propagated through
time. By analogy with the Gaussian case, we consider an initial
wave function having the same functional form as the propagator, namely
\begin{equation}
\psi_0(V) = \frac{\sqrt{V V_0}}{\mathcal{N}_\nu}
\exp\left(-\frac{V^2+V_0^2}{4\sigma^2}\right)
I_\nu\left(\frac{VV_0}{2\sigma^2}+ikV\right),
\end{equation}
whose normalization is found, still using Weber second integral, to be
given by
\begin{equation}
\mathcal{N}_\nu^2 = \sigma^2V_0 \exp \left(-\frac{V_0^2}{4\sigma^2} -
\sigma^2k^2\right)
I_\nu\left(\frac{V_0^2}{4\sigma^2}+\sigma^2k^2\right);
\end{equation}
this choice reduces to Eq.~\eqref{psi0K0} for $\nu \to 1/2$.

In the large volume limit $V, V_0 \gg 1$, we can use the asymptotic
expansion of $I_\nu(x)$,
\begin{equation}\label{asymp}
I_\nu(x) \sim \frac{\ex^x}{\sqrt{2\pi x}} -
\frac{\ex^{-x+i\pi\left(\nu-\frac12\right)}}{\sqrt{2\pi x}},
\end{equation}
(with corrections of order $x^{-1}$ for each term) to obtain
\begin{equation}
\begin{split}
\psi(V) \underrel{V\to\infty}{\approx} &
\frac{\ex^{i\mu}}{\sqrt{\sqrt{2\pi}\sigma}} \Bigg\{
\exp\left[-\frac{(V+V_0)^2}{4\sigma^2}-ikV\right]\\
&-\exp\left[-\frac{(V-V_0)^2}{4\sigma^2} + ikV +
i\pi\left(\nu-\frac12\right)\right] \Bigg\},
\end{split}
\end{equation}
where $\mu$ is a constant phase given by 
$$
\frac{V_0}{2\sigma^2}+ik = 
\sqrt{\displaystyle\frac{V_0^2}{4\sigma^4}+k^2} \ex^{i\mu}.
$$
The asymptotic expansion above shows that our choice of wave packet
reduces to a Gaussian when computed far from the boundary. In addition,
we can again use Weber second integral to calculate explicitly the solution
by applying the propagator to the initial wave function, namely
$$
\psi(V, \eta) = \int_0^\infty\mathrm{d}V^\prime G(\eta, V,
V^\prime)\psi_0(V^\prime).
$$
This yields
\begin{equation}\label{psiVeta}
\begin{split}
\psi(V, \eta) = & \frac{\sqrt{V V_0}}{\mathcal{N}_\nu(\sigma_\eta/\sigma)}
\exp\left(-ik^2\eta-\frac{V^2+V^2_\eta}{4\sigma\sigma_\eta}\right)\\
& \times I_\nu\left(\frac{V V_\eta}{2\sigma\sigma_\eta}+ikV\right),
\end{split}
\end{equation}  
where $V_\eta = V_0 + 2k\eta$ and $\sigma_\eta = \sigma+i\eta/\sigma$ are
the same parameters used in the free particle case. As below
Eq.~\eqref{propK}, the limit $\Ka\to 0$ reproduces the solution
\eqref{sinh}, up to an irrelevant phase.

Far from the boundary, this solution reduces to a simple Gaussian packet
traveling with speed $2k$. On the other hand, if the packet travels
towards the boundary, $V_\eta$ eventually vanishes and consequently
the modified Bessel function argument, whose real part reads
$$\Re\ex \left(\frac{VV_\eta}{2\sigma\sigma_\eta}+ikV\right) =
 \frac{VV_\eta}{2\vert\sigma_\eta\vert^2},$$
also vanishes. At this stage, the asymptotic expansion is clearly not
valid so the wave function cannot be approximated by a Gaussian packet.
Nonetheless, $-V_\eta$ subsequently again increases monotonically,
so that, given enough time, the wave function again behaves as a Gaussian
wave packet traveling away from the boundary. This happens because as the
sign of the real part of the argument changes to negative, the asymptotic
expansion of the modified Bessel function~\eqref{asymp} becomes dominated
by the second term.

With the help of the actual wave function solution \eqref{psiVeta}, it is
possible to estimate the average value of the relevant variable $\hat V$,
namely
$$
\langle \hat{V} \rangle = \int_0^\infty |\psi(V,\eta)|^2 \dd V
$$
and $\hat{\pi}_V$,
$$
\langle \hat{\pi}_V \rangle = - i \int_0^\infty \psi^*(V,\eta) \partial_V \psi(V,\eta) \dd V.
$$
Figure \ref{V2eta} shows $\langle \hat{V}\rangle$ as a function of the
conformal time and compares with the semiclassical trajectory. It is
clear from this figure that although using $\sqrt{\langle \hat{V}^2
\rangle}$ may be questionable, it provides a reasonable approximation to
$\langle \hat{V}\rangle$ almost at all times. This is due to the fact
that we considered a very peaked Gaussian state for large negative times,
and although the variance increases with time after the bounce, the
difference remains small because the growing variance is just compensated
by the simultaneous shift of the wave packet to larger and larger values
of $V$: for large values of $\eta$, we have  $\langle \hat{V}\rangle
\propto \sigma_\eta \propto \eta$.

The relevant phase space trajectory is illustrated on Fig.~\ref{pvpv},
showing again that the semiclassical approximation is a valid one,
especially if one is interested in the asymptotic (large time) behaviors.
The solution \eqref{vtpt} is symmetric, contrary to the mean value case.
This stems from the fact that the variance of the wave packet has a
nonsymmetric evolution in time (Fig.~\ref{bounce2}).

\bibliography{references}

\end{document}